\documentclass[reprint,aps,prd,superscriptaddress,preprintnumbers,nofootinbib]{revtex4-1}

\usepackage{amsfonts}
\usepackage{amsmath}
\usepackage{amsthm}
\usepackage{amssymb}
\usepackage{array}
\usepackage{dcolumn}
\usepackage[svgnames]{xcolor}
\usepackage[hidelinks]{hyperref}
\hypersetup{
    colorlinks=true,
    linkcolor=NavyBlue,
    filecolor=NavyBlue,
    urlcolor=NavyBlue,
    citecolor=NavyBlue,
    }

\usepackage{graphics}
\usepackage{graphicx}
\usepackage{adjustbox}
\usepackage{breqn}

\def\be{\begin{equation}}
\def\ee{\end{equation}}
\def\bea{\begin{eqnarray}}
\def\eea{\end{eqnarray}}

\def\gev{\, {\rm GeV}}
\def\mev{\, {\rm MeV}}
\def\kev{\, {\rm keV}}

\def\yr{\, {\rm yr}}

\newcommand{\cm}{\rm cm}

\makeatletter
\let\cat@comma@active\@empty
\makeatother

\begin{document}

\hfill \preprint{MI-HET-758}

\title{Cosmic-ray Upscattered Inelastic Dark Matter}

\author{Nicole F.~Bell}
\email{n.bell@unimelb.edu.au}
\affiliation{ARC Centre of Excellence for Dark Matter Particle Physics, \\
School of Physics, The University of Melbourne, Victoria 3010, Australia}

\author{James B.~Dent} 
\email{jbdent@shsu.edu}
\affiliation{Department of Physics, Sam Houston State~ University, Huntsville, Texas 77341,~ USA}

\author{Bhaskar Dutta}
\email{dutta@physics.tamu.edu}

\author{Sumit~Ghosh}
\email{ghosh@tamu.edu}
\affiliation{Mitchell Institute for Fundamental Physics and Astronomy, \\
Department of Physics and Astronomy, Texas A\&M University, College Station, Texas 77843,USA}

\author{Jason Kumar}
\email{jkumar@hawaii.edu}
\affiliation{Department of Physics and Astronomy, University of Hawaii, Honolulu, Hawaii 96822, USA}

\author{Jayden L.~Newstead}
\email{jnewstead@unimelb.edu.au}
\affiliation{ARC Centre of Excellence for Dark Matter Particle Physics, \\
School of Physics, The University of Melbourne, Victoria 3010, Australia}

\author{Ian M.~Shoemaker}
\affiliation{Center for Neutrino Physics, Department of Physics, Virginia Tech University, Blacksburg, Virginia 24601, USA}

\begin{abstract}
Light non-relativistic components of the galactic dark matter halo elude direct detection constraints because they lack the kinetic energy to create an observable recoil. However, cosmic-rays can upscatter dark matter to significant energies, giving direct detection experiments access to previously unreachable regions of parameter-space at very low dark matter mass. In this work we extend the cosmic-ray dark matter formalism to models of inelastic dark matter and show that previously inaccessible regions of the mass-splitting parameter space can be probed. Conventional direct detection of non-relativistic halo dark matter is limited to mass splittings of $\delta\sim10~\mathrm{keV}$ and is highly mass dependent. We find that including the effect of cosmic-ray upscattering can extend the reach to mass splittings of $\delta\sim100~\mathrm{MeV}$ and maintain that reach at much lower dark matter mass.
\end{abstract}

\maketitle

\section{Introduction} \label{sec:introduction}

Until recent years low-mass dark matter (DM) was relatively unconstrained by direct detection experiments.  The difficulty low-mass DM presents is that the recoil energy deposited is proportional to the DM mass, typically falling below the detector threshold for masses less than a few GeV. While low-threshold detector technologies have made advances in recent years, new analysis strategies have lead the field in constraining low-mass DM~\cite{Essig:2011nj, Essig:2012yx, Graham:2012su, An:2014twa, Essig:2015cda, Hochberg:2015pha, Derenzo:2016fse, Bloch:2016sjj, Hochberg:2016ntt, Hochberg:2016ajh, Kouvaris:2016afs, Essig:2017kqs, Budnik:2017sbu, Bunting:2017net, Knapen:2017ekk, Hochberg:2017wce, Hertel:2018aal, Dolan:2017xbu, Bringmann:2018cvk, Emken:2019tni, Essig:2019kfe, Ema:2018bih, Bell:2019egg, Trickle:2019ovy,Trickle:2019nya, Griffin:2019mvc, Baxter:2019pnz, Kurinsky:2019pgb, Catena:2019gfa, Griffin:2020lgd, Flambaum:2020xxo, Bell:2021zkr}. Two particularly useful strategies, which have been the subject of several recent studies, are the Migdal effect~\cite{Migdal:1941, Vergados:2004bm, Bernabei:2007jz, Ibe:2017yqa, Dolan:2017xbu, Bell:2019egg, Essig:2019xkx, Liu:2020pat, GrillidiCortona:2020owp, Dey:2020sai, Bell:2021zkr} and cosmic-ray boosted dark matter (CRDM)~\cite{Bringmann:2018cvk, Dent:2019krz, Guo:2020oum}.  
These studies have all focused on elastic nuclear scattering.  However, inelastic DM scattering is a generic feature of many classes of DM models~\cite{TuckerSmith:2001hy, TuckerSmith:2004jv, Finkbeiner:2007kk, Arina:2007tm, Chang:2008gd, Cui:2009xq, Fox:2010bu, Lin:2010sb, DeSimone:2010tf, An:2011uq, Pospelov:2013nea,Finkbeiner:2014sja, Dienes:2014via, Barello:2014uda, Bramante:2016rdh,Bell:2018pkk, Jordan:2018gcd, Dutta:2019fxn,  Bell:2020bes}.  Here we explore the prospects for inelastic DM detection within the CRDM paradigm. 

In the CRDM paradigm, rather than finding a channel through which small energy depositions can be detected (e.g. Migdal electrons), one instead finds a population of fast moving DM which can yield larger energy deposition.  When energetic cosmic rays (mostly protons) scatter off non-relativistic DM particles in the halo, they can produce a small population of relativistic DM.  If these relativistic DM particles scatter at a direct detection experiment, then the deposited energy can be well above threshold.  This relativistic population can thus provide the leading channel at direct detection experiments.

Although non-relativistic inelastic DM scattering has been studied in-depth in the context of the DAMA excess, inelastic scattering is in fact a generic feature of some classes of DM models.  As an illustrative example, we can consider DM which couples to the Standard Model (SM) by exchange of a dark photon.  The DM vector current can only be non-vanishing if the DM is a complex degree of freedom.  But if the continuous symmetries under which the DM is charged are all spontaneously broken, then the DM generically splits into two real degrees of freedom, and the vector current is necessarily off-diagonal, mediating inelastic scattering. 

Previous model building efforts of inelastic DM have focused on small mass splittings, motivated by a desire to either explain an experimental anomaly or to stay in contact with experimentally accessible signals. More generally, there is no reason to presuppose that the mass splitting be $\mathcal{O}$(keV). In the example given above, the mass splitting need only be small relative to the symmetry breaking scale and could easily be $\mathcal{O}$(MeV-GeV). Such large mass splittings are inaccessible to non-relativistic direct detection experiments and have only been probed in collider experiments~\cite{Jordan:2018gcd,Batell:2021ooj}. 

For CRDM, the initial inelastic upscattering process can have a much larger center-of-mass energy, dictated by the cosmic-ray energies available in the interstellar medium. As a result, much larger mass splittings are accessible in this scenario as compared to the standard nuclear recoil case.  Given the long path-length from cosmic-ray upscatter to the detector, we consider two cases: one in which all upscattered particles reach the Earth before decaying where they exothermically scatter in a detector, and one in which all upscattered particles decay before reaching the detector, where they endothermically scatter. 

The plan of this paper is as follows: in Section~\ref{sec:upscattering}, we derive the 
energy spectrum of CR-upscattered inelastic DM (CRiDM). In Section \ref{sec:RecoilSpectrum} we present the  recoil spectrum arising from the inelastic scattering of CRiDM, and comment on the distinguishability of the scenarios under consideration. In Section \ref{sec:Results}, we describe the bound on CRiDM which are placed by XENON1T. Lastly, in Section \ref{sec:conclusion}, we conclude with a discussion of our results and future avenues.

\section{Cosmic-ray upscattering of inelastic DM}
\label{sec:upscattering}

The direct detection of DM relies on a non-zero cross section for the DM scattering on nucleons or electrons. Consequently, there is also the possibility that DM can first be upscattered by cosmic-rays before it reaches the detector~\cite{Bringmann:2018cvk}. Light DM candidates (below a GeV) can be upscattered to relativistic energies, making their recoils visible to experiments that were previously insensitive to them, or conversely, one could see their impact on the cosmic-ray spectrum itself~\cite{Cappiello:2018hsu}. Previous analyses have explored CRDM in the context of simplified models~\cite{Dent:2019krz,Bondarenko:2019vrb}, scattering on electrons~\cite{Ema:2018bih}, scattering in neutrino detectors~\cite{Cappiello:2019qsw} and inelastic hadronic scattering~\cite{Guo:2020oum}. In this work we consider the effect of inelastic scattering due to the DM candidate which couples to nucleons.

Since we will consider processes in which the center-of-mass energy may be much larger than the mass of the mediating particle, it will be necessary to provide a model for DM-SM interactions beyond the contact approximation.  For simplicity, we assume that dark sector particles are two Majorana fermions, $\chi_{1,2}$  ($m_{\chi_2} - m_{\chi_1} \equiv \delta > 0$), which couple to a spin-1 particle ($A'$) through an interaction $g_\chi A'_\mu (\bar \chi_2 \gamma^\mu \chi_1 - \bar \chi_1 \gamma^\mu \chi_2)$.  $A'$ also couples to nucleons through an interaction $g_N A'_\mu \bar n \gamma^\mu n$.  In particular, we consider the case in which $A'$ couples to protons and neutrons with equal strength.

Note that there are some important consistency conditions associated with this effective interaction, in order to ensure that it arises from a consistent theory.  For example, if the coupling $g_N$ remains fixed, then in the limit $m_{A'} \rightarrow 0$ the gauge symmetry is unbroken, and one must have $\delta \rightarrow 0$ as a result of gauge-invariance.  More generally, in order for our tree-level calculation of the cross section to be consistent, one should require $g \lesssim 1$, and $\delta \lesssim m_\chi / g$.  The latter condition ensures that the Yukawa couplings which generate the mass splitting are also perturbative.  In our subsequent analysis, we will focus on regions of parameter space where these constraints are satisfied.

\begin{figure}[hbt]
    \includegraphics[width=0.85\columnwidth]{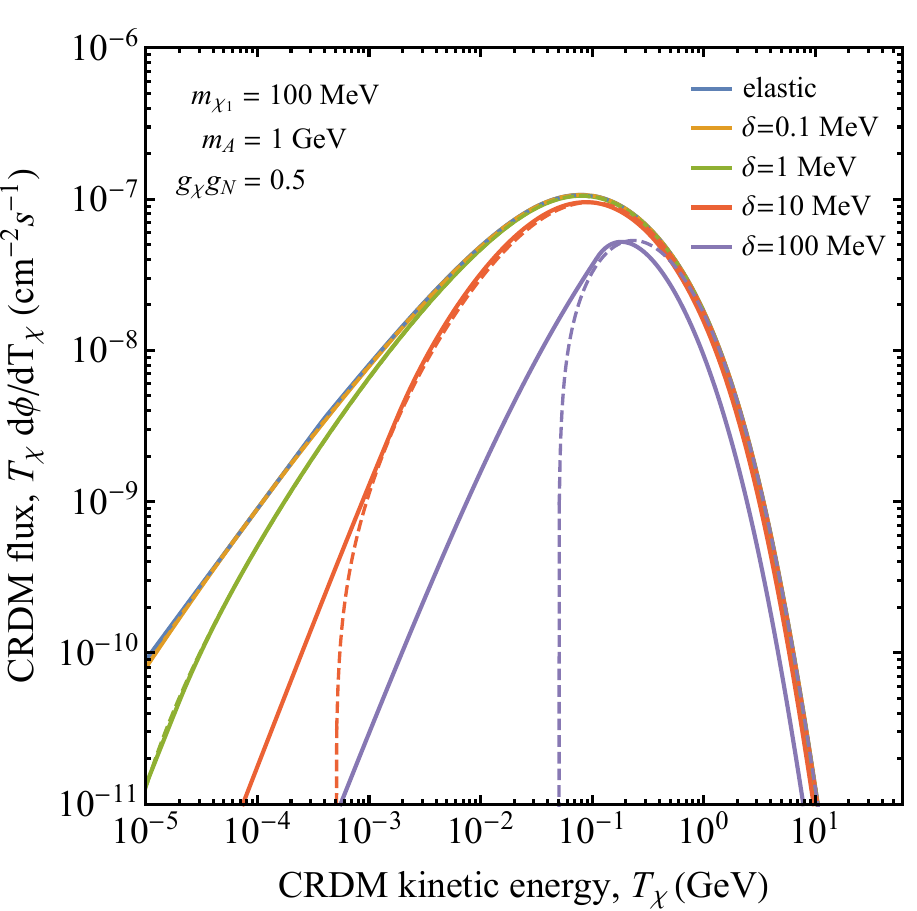}
    \includegraphics[width=0.85\columnwidth]{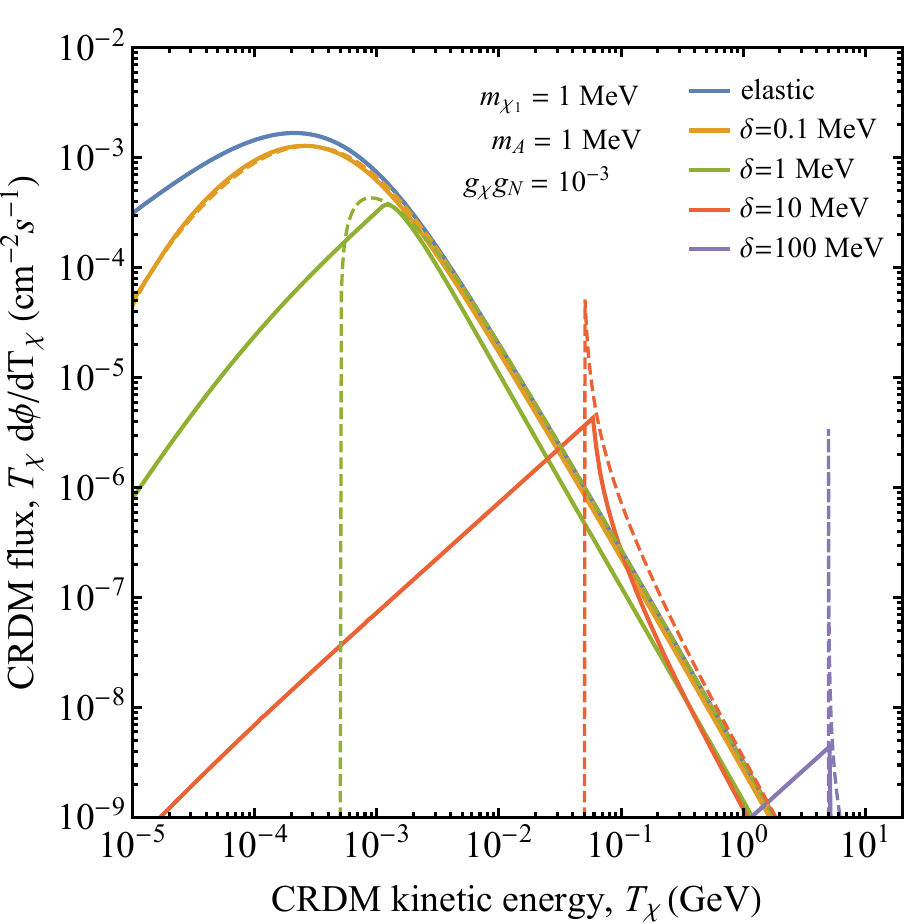}
    \caption{Sample spectra of dark matter after upscattering by cosmic rays ($\chi_2$, dashed) and after subsequently decaying  ($\chi_1$, solid). The approximate non-relativistic total cross sections these couplings correspond to is: $\tilde{\sigma}_0=10^{-31}$cm$^2$ and $\tilde{\sigma}_0=5\times10^{-30}$cm$^2$, for the top and bottom respectively.}
    \label{fig:flux}
\end{figure}

The double-differential rate of cosmic-rays scattering on DM within an infinitesimal volume element is
\bea
\frac{d^2\Gamma}{dT_idT_{\chi_2}} = \frac{\rho_\chi}{m_{\chi_1}} \frac{d\sigma_{\chi i}} {dT_{\chi_2}} \frac{d\Phi^{\rm{LIS}}_i}{dT_i} dV ,
\eea
where $\rho_\chi = 0.3 \gev / \cm^3$ is the local DM density, and $d\Phi^{\rm{LIS}}_i/dT_i$ is the local interstellar flux of the $i$th species of incident cosmic-rays (here we include contributions from protons and helium only, with the spectra taken from~\cite{Boschini:2017fxq}).
$T_i$ is the incoming CR kinetic energy and $T_{\chi_2}$ is the outgoing DM kinetic energy.
$\sigma_{\chi i} (T_i, T_{\chi_2})$ is the cross section for scattering of DM with the $i$th cosmic ray species. The total upscattered DM flux at Earth is obtained by integrating this over the relevant volume and cosmic-ray spectrum, 
\bea
\frac{d\Phi_{\chi_2}}{dT_{\chi_2}} &=& \int_V\!\frac{dV}{4\pi d^2}\int_{T_i^{\rm{min}}}^{T_i^{\rm{max}}}\!\!\!\!dT_{i} \,\,\frac{d^{3} \Gamma}
{dT_idT_{\chi_2} dV } ,\\
&=& D_{\rm eff}\,\frac{\rho_\chi}{m_{\chi_1}}\,  \int_{T_i^{\rm{min}}}^{T_i^{\rm{max}}}\!\!\!\!
dT_{i}\,\frac{d\sigma_{\chi i}}{dT_{\chi_2}}\,\frac{d\Phi^{\rm{LIS}}_i}{dT_i} ,
\eea
where 
$D_{{\rm eff}}$ is an effective diffusion zone parameter. The limits of the energy integral are given by:
\begin{eqnarray}
&& T_i^{\text{max/min}}  = \frac{T_{\chi_2} - 2 m_i+\delta}{2} \nonumber\\
&& \pm \frac{1}{2}\left[\frac{T_{\chi_2} (2 \delta +2 m_{\chi_1}+T_{\chi_2}) \left(4 m_i^2+2 m_{\chi_1} T_{\chi_2} -\delta ^2\right)}{\left(2 m_{\chi_1} T_{\chi_2}-\delta ^2\right)}\right]^{\frac{1}{2}} , \nonumber\\
\end{eqnarray}
which are the maximum and minimum kinetic energy of the incoming cosmic ray, such that it is kinematically possible for the outgoing $\chi_2$ to have kinetic energy $T_{\chi_2}.$

To account for the variation in the DM density throughout the diffusion zone, within which the cosmic-ray flux is assumed to be constant, an effective diffusion zone parameter $D_{{\rm eff}}$ is found by integrating over the NFW profile. These assumptions give rise to some uncertainty in the value of $D_{{\rm eff}}$, which can be minimized by following Ref.~\cite{Bringmann:2018cvk} and conservatively considering a diffusion zone of only 1 kpc, corresponding to $D_{{\rm eff}} = 0.997$ kpc. This choice only modestly reduces the sensitivity of the analysis. 

The differential cross section is:
\begin{widetext}
\bea
\label{eq:diffCS}
\frac{d\sigma_{\chi i}}{dT_{\chi_2}} &=& g_\chi^2 g_N^2 A_i^2 F^2(q^2)\nonumber\\
 & & \times\frac{4 m_\chi (m_i + T_i)^2 - 
       2 ((m_i + m_{\chi_1})^2 + 2 m_{\chi_1} T_i) T_\chi + 2 m_{\chi_1} T_{\chi_2}^2 - 
       4 m_{\chi_1} (m_i + T_i) \delta + (m_{\chi_1} - T_{\chi_2}) \delta^2}{ 2\pi T_i (2 m_i + T_i) (m_A^2 + 2 m_{\chi_1} T_{\chi_2} - \delta^2)^2 } ,
\eea
\end{widetext}
where $A_i$ is the atomic number of the $i$th cosmic-ray species and $F^2(q^2)$ is the form factor. For cosmic-ray$-$DM scattering we follow \cite{Bringmann:2018cvk} and take the hadronic form factor to be of the dipole form with $\Lambda_p = 770 \mev$ and $\Lambda_{\mathrm{He}} = 410 \mev$.

We assume that the lighter particle, $\chi_1$, is the dominant constituent of the galactic DM halo, while the heavier exited state, $\chi_2$, only makes up a negligible fraction. To be satisfied, this assumption requires a sufficiently small upscattering rate and/or a sufficiently short lifetime of the $\chi_2$. However, if the lifetime of the $\chi_2$ is greater than a few years, the flux of CRiDM could be a mixture of these two states. 
We therefore consider two scenarios: one in which the DM decays before reaching the Earth and then endothermically scatters on a target nucleus in the detector, and a second in which the $\chi_2$ reaches the Earth and exothermically scatters on a nucleus in the detector target.
These two scenarios represent the extremal limits, all finite lifetime scenarios will lie between these two cases.

In Figure \ref{fig:flux}, we plot the differential flux of $\chi_2$ (dashed lines) produced by cosmic ray up-scattering of the ambient dark matter distribution, for either $m_{\chi_1} = 100~\mev, 1~\mev$ (upper and lower panels, respectively), and $\delta = 1, 10, 100~\mev$ (orange, green, and red lines, respectively). Note that, for $\delta > 0$, there is always a minimum kinetic energy for upscattered $\chi_2$.  This reflects the fact that, in center-of-mass frame, we must have $|\vec{p}_{\chi_2}| < |\vec{p}_{\chi_1}|$, implying that in the frame of the Earth, $\chi_2$ must be forward-moving. When the mediator mass is small, the spectra for $\delta = 10^{-2}$ GeV and $10^{-1}$ GeV exhibit sharp peaks. This is due to an enhancement in the cross section for outgoing dark matter energies in the vicinity of $T_\chi = \frac{\delta^2-m_A^2}{2m_\chi}$, where the propagator term of Eq.~\ref{eq:diffCS} vanishes. While this point is not kinematically accessible, when the mediator mass is small compared to $\delta$ large values of the incoming cosmic-ray energy can get arbitrarily close.

\subsection{Decay of excited DM}
\label{sec:decay}

The heavier dark sector particle can decay to the lighter particle by the emission of a photon.  This process will affect the energy spectrum of the relativistic dark matter component, and can potentially lead to an observable gamma-ray signature.

\begin{figure}[tbh]
    \centering
    \includegraphics[width=0.9\columnwidth]{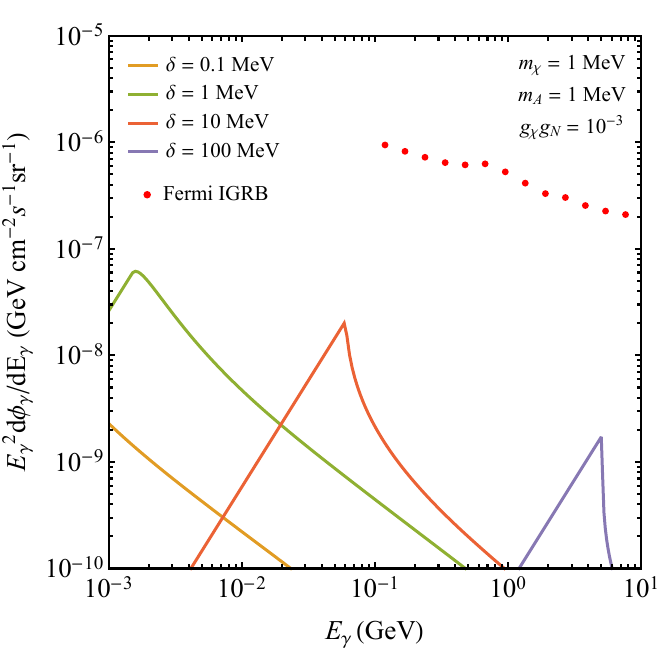} \\
    \caption{Sample spectra of photons resulting for the decay $\chi_2 \rightarrow \chi_1$ for various values of the mass splitting.}
    \label{fig:photons}
\end{figure}

Let us first consider the process $\chi_2 \rightarrow \chi_1 \gamma$, where we assume that the angular distribution is isotropic in the rest frame of the parent particle.  This process cannot proceed through a vector current interaction, as a result of gauge-invariance, but may proceed through a magnetic dipole interaction ($(1/\Lambda) \bar \chi_2 \sigma^{\mu \nu} \chi_1 F_{\mu \nu}$). We can then find the CRiDM and photon spectra using the results of~\cite{Boddy:2016hbp}, which considered this decay process in the context of indirect detection.

In the rest frame of $\chi_2$, the energies of the lighter DM particle ($E_{\chi_1}$) and photon ($E_*$) are given by 
\bea
E_{\chi_1}&=& \frac{m_{\chi_2}^2 + m_{\chi_1}^2}{2m_{\chi_2}},\\
E_* &=& \frac{m_{\chi_2}^2 - m_{\chi_1}^2}{2m_{\chi_2}},\\
\nonumber
\eea 
respectively. 

The post-decay CRiDM flux in the detector frame is then given by,
\bea
\frac{d\Phi_{\chi_1}}{dT_{\chi_1}} & & =  \int_{T_{\chi 2}^{\text{min}}}^{T_{\chi 2}^{\text{max}}} dT_{\chi_2} \frac{d\Phi_{\chi_2}}{dT_{\chi_2}}\nonumber\\
&  &  \times\frac{m_{\chi_2}^2}{(m_{\chi_1}+m_{\chi_2})\delta\sqrt{T_{\chi_2}(2 m_{\chi_1}+2\delta+T_{\chi 2})}}.
\label{eqn:Chi1Spectrum}
\eea
Assuming that all of the $\chi_2$ decay produced by CR upscattering decay back to $\chi_1$, the differential spectrum of this $\chi_1$ population is also shown in Fig.~\ref{fig:flux} (solid lines).

In the frame of the detector, the photon spectrum is then given by~\cite{Boddy:2016hbp}
\bea 
\frac{d\Phi_\gamma}{dE_\gamma} &=& \int_{\frac{m_{\chi_2}}{2}
\left(x + \frac{1}{x} \right)}^\infty
\!\!dE_{\chi_2} \!\!\left[ \frac{d\Phi_{\chi_2}}{dE_{\chi_2}} 
\frac{m_{\chi_2}}{2E_* \sqrt{E_{\chi_2}^2-m_{\chi_2}^2}} 
\right] ,
\label{eqn:GammaSpectrum}
\eea
where $x \equiv E_\gamma / E_*$, $E_\gamma$ is the photon energy in the Earth frame. The resulting gamma ray flux is shown in Fig.~\ref{fig:photons}, and is generically subdominant to the intergalactic gamma-ray background for the relevant couplings.

\section{Detection of CRiDM}
\label{sec:RecoilSpectrum}
The upscattered $\chi_1$ ($\chi_2$) flux arriving at the Earth can then endothermically (exothermically) scatter in a detector. The differential event rate (per unit detector mass) from the incoming DM flux that can be measured by a terrestrial detector is given by
\be 
\frac{dR}{dE_{T}} =\frac{1}{m_{T}} \int_{T_{\chi}^{{\rm min}}}^{T_{\chi}^{{\rm max}}} dT_{\chi}~\frac{d\Phi_\chi}{dT_\chi}~ \frac{d \sigma_{\chi T}}{dE_{T}} ,
\label{eq:rate}
\ee
where $E_{T}$ is the recoil energy of the target nuclei, with mass $m_T$. In contrast to the elastic case, the integral over the DM energies now includes an upper limit due to the inelastic nature of the scattering. The kinematic limits are given by
\begin{widetext}
\bea
T_{\chi_j}^{\rm min/max} & = & \frac{1}{2}E_T m_T-\frac{\delta}{4}(2m_{\chi_1}+\delta)  - m_{\chi_j}\nonumber\\
& & \pm\frac{\left[E_T(E_T+2m_T)(2E_T m_T+\delta^2)(2E_T m_T+(2m_{\chi_1}+\delta)^2)\right]^{1/2}}{4E_T m_T}
\label{eq:limits}
\eea
where $j=1,2$ corresponds to endothermic and exothermic scattering respectively. The differential cross sections for scattering on a nuclear target with $A$ nucleons are given by
\bea
\frac{d\sigma_{\chi_1 T}}{dE_{T}} & & = g_\chi^2 g_N^2 A^2 F^2\left(E_T\right) \nonumber \\
 & & \times\frac{2 m_T E_T^2-E_T\left(2(m_T+m_{\chi_1})^2+4m_T T_{\chi_1}+2m_\chi \delta+\delta^2\right)+m_T\left(4(m_{\chi_1}+T_{\chi_1})^2-\delta^2\right)}{2\pi\left(m_A^2+2m_T E_T\right)^2(T_{\chi_1}^2+2m_{\chi_1}  T_{\chi_1})} 
 \label{eq:CrossSec1}
\eea
and
\bea
\frac{d\sigma_{\chi_2 T}}{dE_{T}} & & = g_\chi^2 g_N^2 A^2 F^2\left(E_T\right) \nonumber\\
 & & \times\frac{2 m_T E_T^2-E_T\left(2(m_T+m_{\chi_1})^2+4m_T T_{\chi_2}+4m_T\delta-\delta^2\right)+m_T\left(4(m_{\chi_1}+T_{\chi_2})^2+6(m_{\chi_1}+T_{\chi_2})\delta+\delta^2\right)}{2\pi\left(m_A^2+2m_T E_T\right)^2\left(T_{\chi_2}^2+2(m_{\chi_1}+\delta)  T_{\chi_2}\right)}
 \label{eq:CrossSec2}
\eea
\end{widetext}
for endothermic and exothermic scattering, respectively. For DM-nuclear scattering we take the form factor to be of the Helm form~\cite{Helm:1956zz}.
It is conventional to express the DM-proton interaction strength in terms of the total cross section at zero momentum transfer. This quantity cannot capture the physics of the high-energy scattering taking place. However, to make contact with the bounds on the non-relativistic cross section we define a reference cross-section $\tilde{\sigma}_0 \equiv 4 g_\chi^2 g_N^2 \mu_{\chi N}^2 / \pi m_{A}^4$.

An upper bound of $T_\chi = 100$ GeV is placed on the DM kinetic energy, as higher energies do not have a significant enough flux for the XENON1T detector. Higher energies and potentially larger mass splittings may be accessible with large volume neutrino detectors, and we leave such explorations to a future work.

\begin{figure}
    \centering
    \includegraphics[width=0.85\columnwidth]{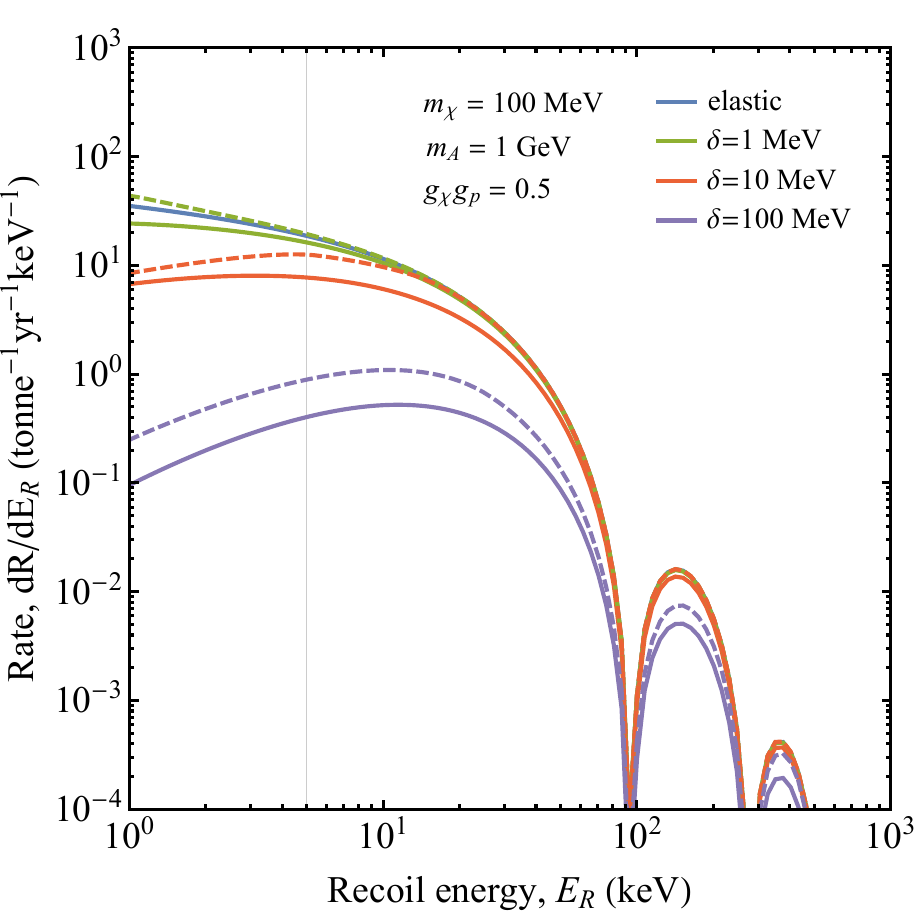}
    \includegraphics[width=0.85\columnwidth]{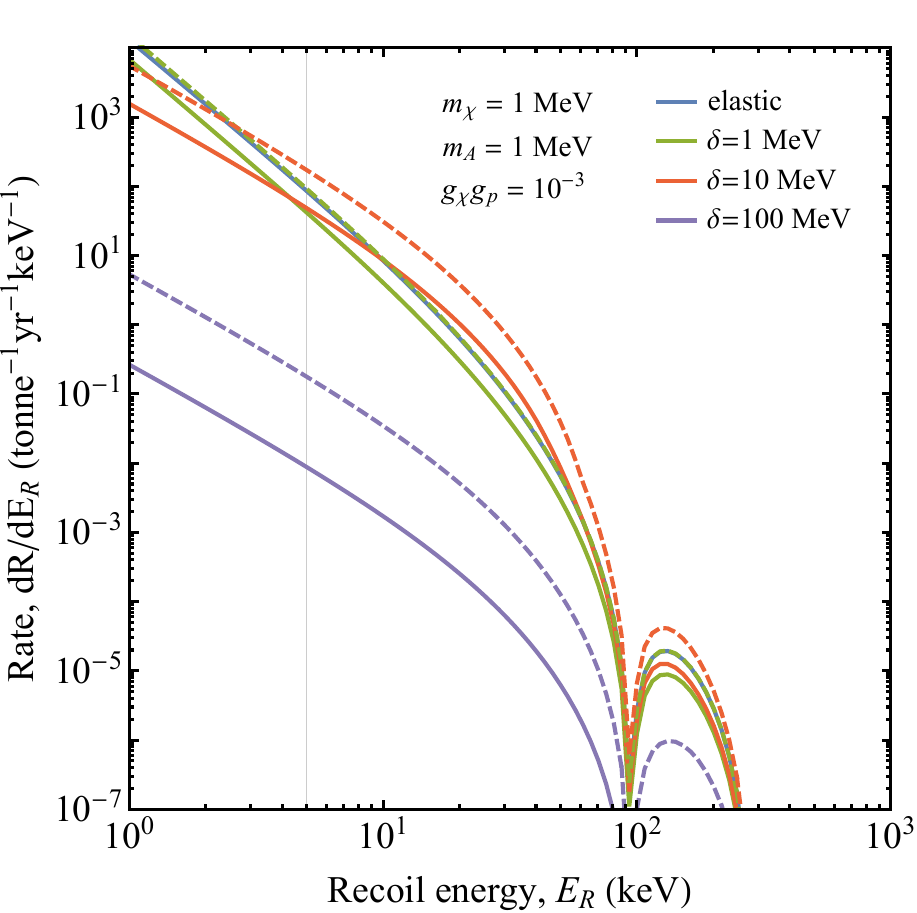}
    \caption{Rates of cosmic-ray dark matter scattering in xenon with a 1 GeV mediator (top) and a 1 MeV mediator (bottom). Endothermic and exothermic scattering are shown in solid and dashed respectively, for various values of the mass splitting (the $\delta=0.1$ MeV is omitted since it is essentially degenerate with the elastic case). The vertical line denotes the approximate threshold of XENON1T. The approximate non-relativistic total cross sections these couplings correspond to is: $\tilde{\sigma}_0=10^{-31}$cm$^2$ and $\tilde{\sigma}_0=5\times10^{-30}$cm$^2$, for the top and bottom respectively.}
    \label{fig:crdm_rate}
\end{figure}

The resulting differential rates for endo- and exothermic scattering are shown in Fig.~\ref{fig:crdm_rate}. The high-energy of the incoming CRiDM has two main effects: large mass splittings of up to 0.1 GeV are accessible, while small mass splittings  $\delta \lesssim 10$ MeV appear degenerate with the elastic case, especially above the detector threshold. In the case of exothermic scattering, sufficiently small mass splittings can slightly enhance the observable rate, but this is not as pronounced as in the non-relativistic case.

\begin{figure}
    \centering
    \includegraphics[width=0.85\columnwidth]{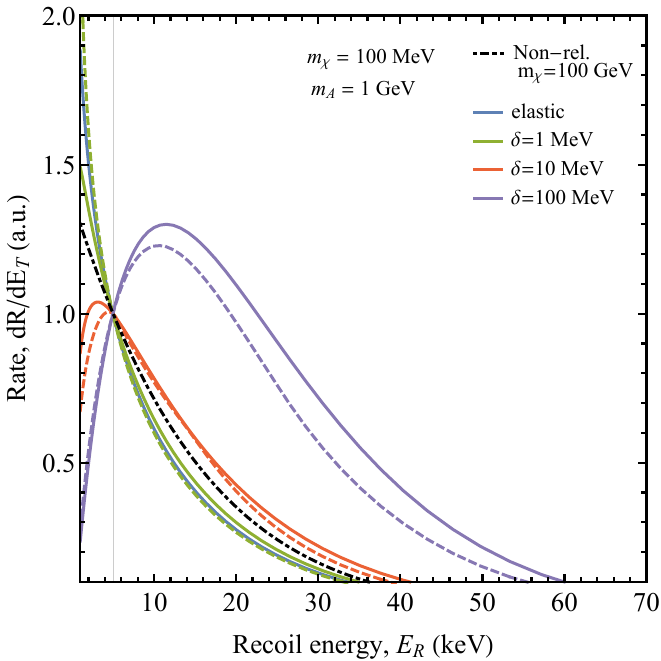}
    \includegraphics[width=0.85\columnwidth]{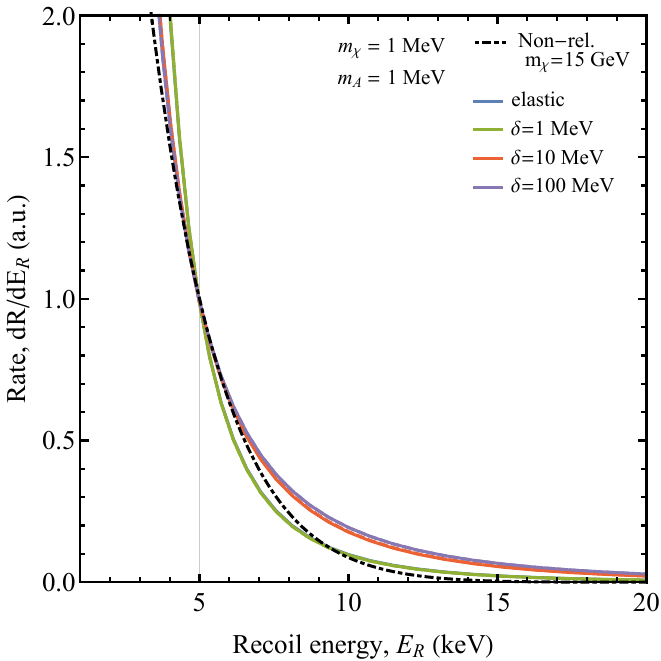}
    \caption{Rates of cosmic-ray dark matter scattering in xenon with a 1 GeV mediator (top) and a 1 MeV mediator (bottom) normalized to the scattering rate at threshold ($E_R=5\kev$). Endothermic and exothermic scattering are shown in solid and dashed respectively, for various values of the mass splitting. For comparison a sample non-relativistic scattering rate with similar energy dependence is shown in black dashed. The vertical line denotes the approximate threshold of XENON1T.}
    \label{fig:rate_scaled}
\end{figure}

To highlight the degeneracy in the event rate we plot the spectra normalized to the elastic rate at $E_T=5\kev$ in Fig.~\ref{fig:rate_scaled}. We chose to plot this on a linear spectrum because it better represents the likelihood of distinguishablity with the low count rates expected of a dark matter signal. With the exception of the $m_{\chi_1}=100\mev$ and $\delta=100\mev$ case, the observable event rate spectra exhibit an approximate degeneracy between variation of the mass splitting and the coupling strength.  Moreover, these scenarios are also approximately degenerate with elastic scattering of non-relativistic heavy dark matter. The degeneracy between mass splitting and coupling strength is present whether there is endothermic or exothermic scattering. 

To explain this similarity, we can consider inelastic DM-nucleus scattering ($\chi_i A \rightarrow \chi_j A$) in the center-of-mass frame.  This scattering process describes both the initial CR-upscatter of DM, as well as inelastic scattering (endothermic or exothermic) of relativistic DM at the detector.  For the model we consider, DM-nucleus scattering is isotropic in the center-of-mass frame.  Thus, given a center-of-mass energy $\sqrt{s}$, assuming a target at rest, the energy spectrum of the outgoing products is determined by the magnitude of the momentum of the outgoing products in the center-of-mass frame, which is given by  $p = [s - (m_{\chi_j} + m_A)^2]/2\sqrt{s}$.  We thus see that if $\delta \ll \sqrt{s} - (m_A + m_\chi)$, then the momentum of the outgoing particle will be largely independent of $\delta$.  In this case, although the normalization of the event rate may depend on $\delta$, the energy spectrum of the outgoing particles will not. 

Thus, if $\delta \ll m_\chi$, we see that the recoil spectrum due to CRiDM is essentially independent of $\delta$, once the energy is well above the threshold for upscattering.  Similarly, since the target detector nucleus is much heavier than the DM, we will have $\delta \ll m_T$ for all scenarios which we consider.  The event rate at the detector will be dominated by events well above threshold, for which again we find that the energy spectrum of the outgoing particles is independent of $\delta$.  We can confirm this from Eq.~\ref{eq:CrossSec1} and Eq.~\ref{eq:CrossSec2}, which show that the differential scattering cross section at the detector is largely independent of $\delta$ if $\delta / m_T \ll 1$.

For the case with $m_{\chi_1}=1\mev$, the CRiDM spectra can be very different at low-energies if $\delta > m_{\chi_1}$. But the kinematics of light DM scattering at the detector requires $T_\chi\gtrsim0.1\gev$ and so only the region where the fluxes are equivalent is probed.

There are several avenues to pursue in order to break the degeneracy in the recoil spectrum. Firstly, at low recoil energies the rate starts to differ and so lower detector thresholds could aid in discrimination - especially for heavier DM. A potential way to discriminate the endo/exothermic scenarios would be to observe the gamma-ray flux arising from the decay process.  However the flux is very small and would be subject to astrophysical backgrounds. The degeneracy with the heavy non-relativistic DM rate could be broken by exploring the kinematic endpoint, which would be much larger for CRDM. While the flux and rate fall dramatically with energy, large volume neutrino detectors may still be sensitive to a CRDM signal.

Note that the exothermic scattering ($\chi_2N\rightarrow\chi_1N$) of DM at the detector is only possible if the lifetime of the excited state is large enough for the particle to travel from the location of cosmic-ray upscattering to the detector without decaying.  Given the size of the effective diffusion zone and the typical energies of the upscattered DM, one finds that the lifetime must be $\gtrsim {\cal O}(10~\yr)$ in order for exothermic scattering to occur at the detector.  We have considered the case in which decay of the excited state occurs by the process $\chi_2 \rightarrow \chi_1 \gamma$, mediated by an inelastic magnetic dipole operator.  Although this is a simple two-body final state, there is no natural scale for the magnetic dipole operator, since the DM is electrically neutral.  As such, there is no reason why the partial decay width to this channel cannot be sufficiently small.  

The decay process $\chi_2 \rightarrow \chi_1 \gamma \gamma \gamma$ can be mediated directly by the vector current interaction, with the $A'$ and three photons coupling to a box diagram.  For small masses and mass splittings, the size of this decay rate can be estimated from chiral perturbation theory~\cite{Dienes:2017ylr}, yielding  $\Gamma_{\chi_2 \rightarrow \chi_1 \gamma \gamma \gamma} \propto \alpha_{em}^3 \delta^{13} / m_\pi^4 f_\pi^4 \Lambda^4$, where $\Lambda \sim max[\delta, m_{A}]$.  This partial decay width depends very strongly on $\delta$; for large enough $\delta$, $\chi_2$ will decay before reaching the detector.  In particular, we find that for $m_{A} \sim 1\gev$, exothermic scattering at the detector is only possible for $\delta \lesssim 25\mev$.  Similarly, if $m_A \lesssim \delta$, then exothermic scattering at the detector is only possible for $\delta \lesssim 5\mev$.

\section{Experimental bounds on CRiDM}
\label{sec:Results}
\begin{figure}[tbh]
    \centering
    \includegraphics[width=0.95\columnwidth]{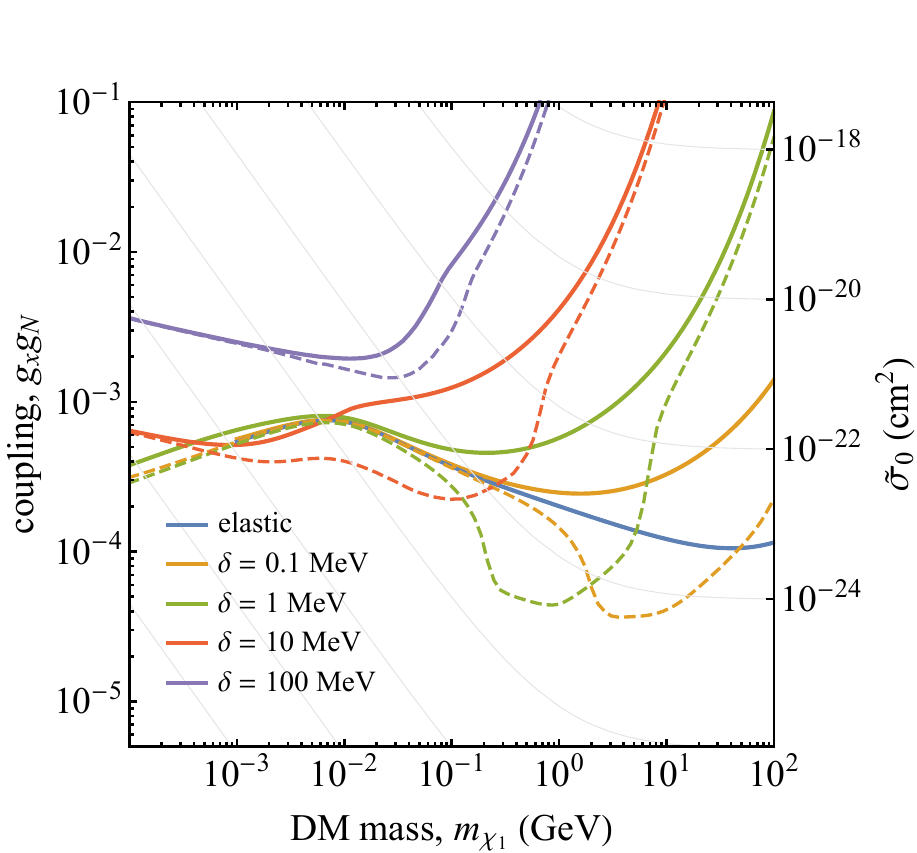}\\
    \includegraphics[width=0.95\columnwidth]{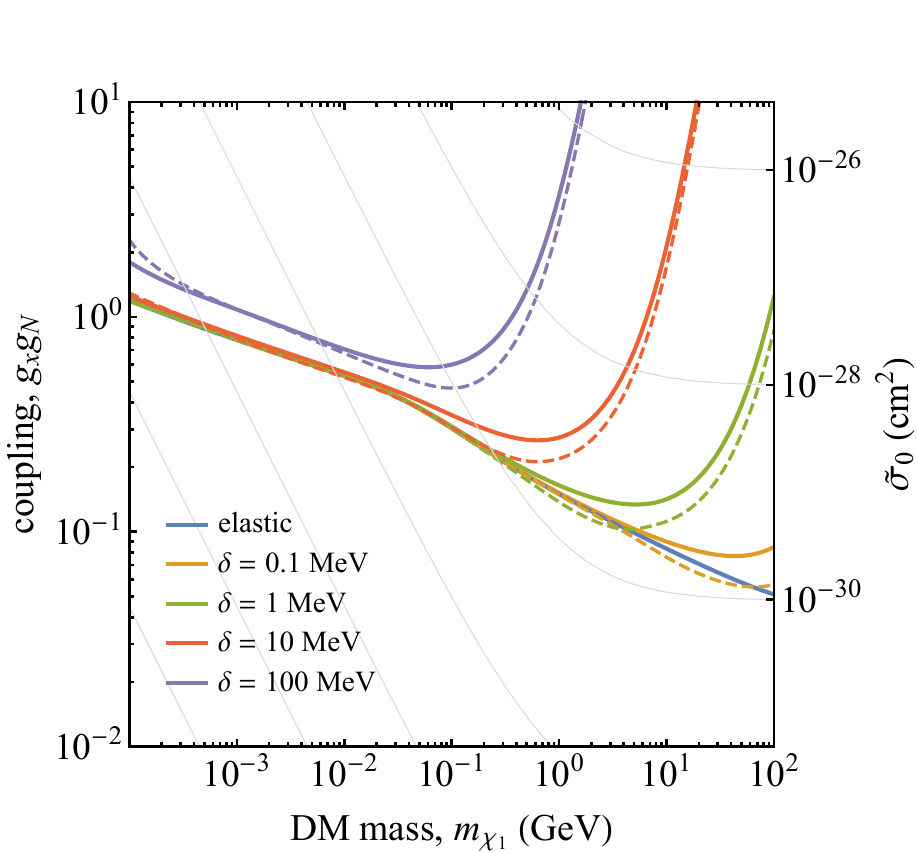}\\
    \caption{Bounds on inelastic CRDM from XENON1T for mediator mass of 1 MeV (top) and 1 GeV (bottom) for various values of the mass splitting where endothermic (exothermic) scattering in the detector is shown in solid (dashed). The light grey contours denote the equivalent non-relativistic cross section for the given mass and coupling (denoted on right axes).}
    \label{fig:bounds}
\end{figure}

At present, the most stringent constraints on CRDM come from the XENON1T experiment. We place bounds on the product of the couplings, $g_\chi g_N$, for two benchmark values for the mediator mass, $m_A=1$~MeV and 1~GeV. The bounds are calculated for a variety of accessible mass splittings across a wide mass range. We compute the upper limits by finding the coupling value that would produce a total of 12 events in a 1 tonne-year exposure (the 90\%CL for additional events given the expected background of 7.36 and observation of 14 events). The total number of events is obtained by integrating the differential rate between recoil energies of 2 to 60 keV, and folding in the energy dependent nuclear-recoil detection efficiency from~\cite{XENON:2018voc}. The resulting bounds are shown in Fig.~\ref{fig:bounds}.

One feature which is immediately apparent is that CR upscattering allows for the direct detection of a relativistic component of dark matter, even if the mass splitting is so large that direct detection of the non-relativistic component would not be kinematically allowed.  Even if inelastic scattering of the non-relativistic component is not allowed on Earth, gravitational infall may allow scattering to proceed in the Sun, leading to gravitational capture~\cite{Nussinov:2009ft,Menon:2009qj,Kumar:2012uh}. In that case, neutrino detectors which look for the neutrinos produced by DM annihilation may be sensitive to these models (see, for example,~\cite{DUNE:2021gbm}).  But CR upscattering allows direct detection of the relativistic component even for mass splittings so large that scattering in the Sun is not allowed.

For the most part, this enhanced sensitivity does not significantly depend on whether or not the upscattered DM decays before reaching the detector, because the kinetic energy of the dark particle incident at the detector is in any case well above the kinematic threshold for inelastic scattering. In general we find that with increased mass splittings the bounds become less stringent. However, for some regions of the parameter space, particularly in the low-mass mediator case, we find that exothermic scattering can significantly increase the sensitivity beyond the elastic scattering case. This effect is due to the kinematics of the exothermic scattering, for a given value of the mass splitting (up to a maximum), there is a mass at which $T_{\chi_2}^{\mathrm{min}}$ (Eq.~\ref{eq:limits}) vanishes. This mass is approximately given by $ E_T m_T / \delta \sim 10^{-3} \left[\mathrm{GeV}\right]^2/\delta$, for a representative recoil energy of $E_T = 10$ keV. Around this mass upscattered particles of a wider range of energies can produce a signal above threshold, enhancing the rate. For a fixed $\delta$, this enhancement in sensitivity disappears for small enough $m_{\chi_2}$; in this regime, most of the kinetic energy released by exothermic scattering is transferred to the outgoing DM particle, not the nucleus.

For the large interaction cross sections under consideration it becomes important to consider the effects of attenuation of the dark matter in the overburden. If the dark matter interacts too strongly it will not be able to penetrate to the depth of the XENON1T detector at LNGS. While a detailed treatment of attenuation due to inelastic scattering is beyond the scope of this work, we estimate the couplings at which the effects of attenuation become important following~\cite{Dent:2019krz}. We solve the energy loss equation for elastic scattering and determine the coupling whereby a dark matter particle arriving at the Earth with $T_\chi=10$ GeV would be attenuated to an energy where the maximum recoil energy it can impart is below the XENON1T threshold (approximately $T_\chi=10^{-3} m_\chi$). Using the elastic scattering cross section to estimate the attenuation should be conservative as inelastic scattering reduces the total cross section.  We note that our treatment of CRDM attenuation differs from that of references \cite{Bringmann:2018cvk} and \cite{Bondarenko:2019vrb} which did not include a nuclear form factor in the attenuation cross section. This simplifies the calculation but artificially enhances scattering at higher momentum transfers. This has the effect of overestimating the affect of attenuation, especially at higher DM masses.

In the non-relativistic case, attenuation causes XENON1T to be blind to DM with nucleon cross sections $\gtrsim 10^{-31}$ cm$^2$. Therefore, for elastic scattering, the XENON1T CRDM signal can provide stronger constraints than the non-relativistic signal for DM masses even above 100 GeV. However, DM masses larger than $\gtrsim 0.1$ GeV will have stronger non-relativistic constraints from small-volume shallow detectors (e.g. the CRESST surface run~\cite{CRESST:2017ues}), rocket-based detectors (e.g.~XQC~\cite{Erickcek:2007jv}) or cosmologically derived bounds (e.g.~\cite{Boddy:2018kfv}). For the inelastic model considered here these constraints are greatly weakened since non-relativistic DM cannot access mass splittings $\gtrsim 0.1$ MeV for $m_\chi \lesssim 40$ GeV. We therefore do not include any non-relativistic constraints as they are not competitive beyond the elastic scattering scenario.

\begin{figure}[tbh]
    \centering
    \includegraphics[width=0.95\columnwidth]{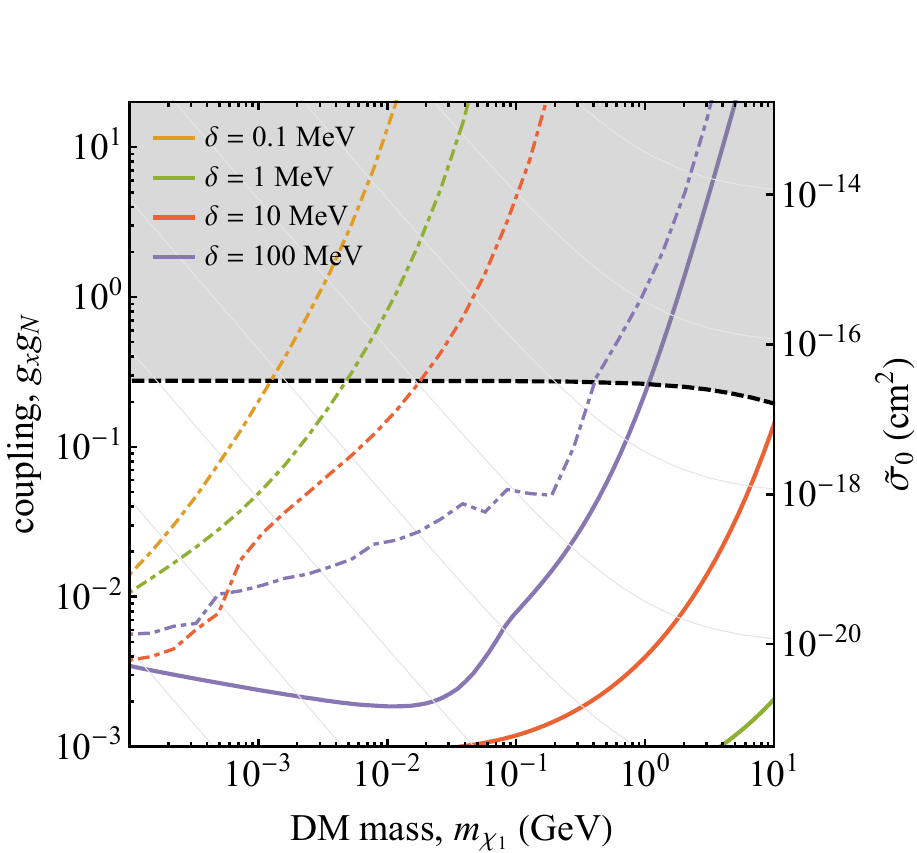}\\
    \caption{Constraints on inelastic CRDM due to endothermic scattering in XENON1T (solid) and the gamma-ray flux from FERMI (dot-dashed), 
    assuming $m_{A'} = 1\mev$. Additionally we show the region where the effect of attenuation can not be ignored and thus XENON1T would be blind to CRDM (gray).}
    \label{fig:fermiLimit}
\end{figure}

A further consideration for the large coupling region is that the decay photons produced in the endothermic scenario may be observable by the FERMI satellite. We set an upper limit where the decay photon flux would exceed 20\% of the FERMI intergalactic gamma-ray background~\cite{Fermi-LAT:2015qzw}. The resulting bound is shown in Fig.~\ref{fig:fermiLimit} for the light mediator case only, since in the heavy mediator case we are not able to constrain perturbative values of the coupling. As expected from the gamma-ray flux derived in section~\ref{sec:decay}, the constraints derived from the decay process are subdominant compared to those from direct detection experiments. However, the gamma-ray flux is not subject to the effects of attenuation and thus the FERMI bounds provide a complementary search channel in this large-coupling region.

Future direct detection experiments with larger exposures and/or lower thresholds such as LZ or SuperCDMS will not greatly improve on the bounds derived in this work. This is because CRDM is not very sensitive to threshold and requires two scattering events, meaning that the bounds on the product of the couplings will scale as 
$ \propto ({\rm exposure})^{-1/4}$.

\section{conclusion} 
\label{sec:conclusion}

In this paper we have extended the CRDM paradigm to include a model of inelastic DM. We have derived constraints on two scenarios, where the DM decays en-route to a terrestrial detector and where the DM exothermically scatters inside the detector. The strongest constraints were found to come from XENON1T, however, in the region of large coupling, attenuation effects become important and constraints on the DM decay from FERMI will be competitive.

The large center-of-mass energy available in CR collisions has allowed us to probe halo DM with much larger mass splittings and much smaller DM masses than can be probed with traditional non-relativistic direct detection.

Interestingly, we find that the nuclear recoil spectrum observed at the detector exhibits degeneracies between the dark matter mass, mass splitting, and scattering cross section.  To break this degeneracy, one could use lower threshold detectors.

\begin{acknowledgments}

The work of BD and SG are supported in part by the DOE Grant No. DE-SC0010813. The work of JK is supported in part by DOE grant DE-SC0010504. JLN and NFB were supported by the Australian Research Council through the ARC Centre of Excellence for Dark Matter Particle Physics, CE200100008. The work of IMS is supported by the U.S. Department of
Energy under the award number DE-SC0020250.

\end{acknowledgments}

\bibliographystyle{apsrev4-1.bst}
\bibliography{cosmic.bib}

\begin{thebibliography}{78}%
\makeatletter
\providecommand \@ifxundefined [1]{%
 \@ifx{#1\undefined}
}%
\providecommand \@ifnum [1]{%
 \ifnum #1\expandafter \@firstoftwo
 \else \expandafter \@secondoftwo
 \fi
}%
\providecommand \@ifx [1]{%
 \ifx #1\expandafter \@firstoftwo
 \else \expandafter \@secondoftwo
 \fi
}%
\providecommand \natexlab [1]{#1}%
\providecommand \enquote  [1]{``#1''}%
\providecommand \bibnamefont  [1]{#1}%
\providecommand \bibfnamefont [1]{#1}%
\providecommand \citenamefont [1]{#1}%
\providecommand \href@noop [0]{\@secondoftwo}%
\providecommand \href [0]{\begingroup \@sanitize@url \@href}%
\providecommand \@href[1]{\@@startlink{#1}\@@href}%
\providecommand \@@href[1]{\endgroup#1\@@endlink}%
\providecommand \@sanitize@url [0]{\catcode `\\12\catcode `\$12\catcode
  `\&12\catcode `\#12\catcode `\^12\catcode `\_12\catcode `\%12\relax}%
\providecommand \@@startlink[1]{}%
\providecommand \@@endlink[0]{}%
\providecommand \url  [0]{\begingroup\@sanitize@url \@url }%
\providecommand \@url [1]{\endgroup\@href {#1}{\urlprefix }}%
\providecommand \urlprefix  [0]{URL }%
\providecommand \Eprint [0]{\href }%
\providecommand \doibase [0]{http://dx.doi.org/}%
\providecommand \selectlanguage [0]{\@gobble}%
\providecommand \bibinfo  [0]{\@secondoftwo}%
\providecommand \bibfield  [0]{\@secondoftwo}%
\providecommand \translation [1]{[#1]}%
\providecommand \BibitemOpen [0]{}%
\providecommand \bibitemStop [0]{}%
\providecommand \bibitemNoStop [0]{.\EOS\space}%
\providecommand \EOS [0]{\spacefactor3000\relax}%
\providecommand \BibitemShut  [1]{\csname bibitem#1\endcsname}%
\let\auto@bib@innerbib\@empty
\bibitem [{\citenamefont {Essig}\ \emph
  {et~al.}(2012{\natexlab{a}})\citenamefont {Essig}, \citenamefont {Mardon},\
  and\ \citenamefont {Volansky}}]{Essig:2011nj}%
  \BibitemOpen
  \bibfield  {author} {\bibinfo {author} {\bibfnamefont {R.}~\bibnamefont
  {Essig}}, \bibinfo {author} {\bibfnamefont {J.}~\bibnamefont {Mardon}}, \
  and\ \bibinfo {author} {\bibfnamefont {T.}~\bibnamefont {Volansky}},\ }\href
  {\doibase 10.1103/PhysRevD.85.076007} {\bibfield  {journal} {\bibinfo
  {journal} {Phys. Rev. D}\ }\textbf {\bibinfo {volume} {85}},\ \bibinfo
  {pages} {076007} (\bibinfo {year} {2012}{\natexlab{a}})},\ \Eprint
  {http://arxiv.org/abs/1108.5383} {arXiv:1108.5383 [hep-ph]} \BibitemShut
  {NoStop}%
\bibitem [{\citenamefont {Essig}\ \emph
  {et~al.}(2012{\natexlab{b}})\citenamefont {Essig}, \citenamefont
  {Manalaysay}, \citenamefont {Mardon}, \citenamefont {Sorensen},\ and\
  \citenamefont {Volansky}}]{Essig:2012yx}%
  \BibitemOpen
  \bibfield  {author} {\bibinfo {author} {\bibfnamefont {R.}~\bibnamefont
  {Essig}}, \bibinfo {author} {\bibfnamefont {A.}~\bibnamefont {Manalaysay}},
  \bibinfo {author} {\bibfnamefont {J.}~\bibnamefont {Mardon}}, \bibinfo
  {author} {\bibfnamefont {P.}~\bibnamefont {Sorensen}}, \ and\ \bibinfo
  {author} {\bibfnamefont {T.}~\bibnamefont {Volansky}},\ }\href {\doibase
  10.1103/PhysRevLett.109.021301} {\bibfield  {journal} {\bibinfo  {journal}
  {Phys. Rev. Lett.}\ }\textbf {\bibinfo {volume} {109}},\ \bibinfo {pages}
  {021301} (\bibinfo {year} {2012}{\natexlab{b}})},\ \Eprint
  {http://arxiv.org/abs/1206.2644} {arXiv:1206.2644 [astro-ph.CO]} \BibitemShut
  {NoStop}%
\bibitem [{\citenamefont {Graham}\ \emph {et~al.}(2012)\citenamefont {Graham},
  \citenamefont {Kaplan}, \citenamefont {Rajendran},\ and\ \citenamefont
  {Walters}}]{Graham:2012su}%
  \BibitemOpen
  \bibfield  {author} {\bibinfo {author} {\bibfnamefont {P.~W.}\ \bibnamefont
  {Graham}}, \bibinfo {author} {\bibfnamefont {D.~E.}\ \bibnamefont {Kaplan}},
  \bibinfo {author} {\bibfnamefont {S.}~\bibnamefont {Rajendran}}, \ and\
  \bibinfo {author} {\bibfnamefont {M.~T.}\ \bibnamefont {Walters}},\ }\href
  {\doibase 10.1016/j.dark.2012.09.001} {\bibfield  {journal} {\bibinfo
  {journal} {Phys. Dark Univ.}\ }\textbf {\bibinfo {volume} {1}},\ \bibinfo
  {pages} {32} (\bibinfo {year} {2012})},\ \Eprint
  {http://arxiv.org/abs/1203.2531} {arXiv:1203.2531 [hep-ph]} \BibitemShut
  {NoStop}%
\bibitem [{\citenamefont {An}\ \emph {et~al.}(2015)\citenamefont {An},
  \citenamefont {Pospelov}, \citenamefont {Pradler},\ and\ \citenamefont
  {Ritz}}]{An:2014twa}%
  \BibitemOpen
  \bibfield  {author} {\bibinfo {author} {\bibfnamefont {H.}~\bibnamefont
  {An}}, \bibinfo {author} {\bibfnamefont {M.}~\bibnamefont {Pospelov}},
  \bibinfo {author} {\bibfnamefont {J.}~\bibnamefont {Pradler}}, \ and\
  \bibinfo {author} {\bibfnamefont {A.}~\bibnamefont {Ritz}},\ }\href {\doibase
  10.1016/j.physletb.2015.06.018} {\bibfield  {journal} {\bibinfo  {journal}
  {Phys. Lett. B}\ }\textbf {\bibinfo {volume} {747}},\ \bibinfo {pages} {331}
  (\bibinfo {year} {2015})},\ \Eprint {http://arxiv.org/abs/1412.8378}
  {arXiv:1412.8378 [hep-ph]} \BibitemShut {NoStop}%
\bibitem [{\citenamefont {Essig}\ \emph {et~al.}(2016)\citenamefont {Essig},
  \citenamefont {Fernandez-Serra}, \citenamefont {Mardon}, \citenamefont
  {Soto}, \citenamefont {Volansky},\ and\ \citenamefont {Yu}}]{Essig:2015cda}%
  \BibitemOpen
  \bibfield  {author} {\bibinfo {author} {\bibfnamefont {R.}~\bibnamefont
  {Essig}}, \bibinfo {author} {\bibfnamefont {M.}~\bibnamefont
  {Fernandez-Serra}}, \bibinfo {author} {\bibfnamefont {J.}~\bibnamefont
  {Mardon}}, \bibinfo {author} {\bibfnamefont {A.}~\bibnamefont {Soto}},
  \bibinfo {author} {\bibfnamefont {T.}~\bibnamefont {Volansky}}, \ and\
  \bibinfo {author} {\bibfnamefont {T.-T.}\ \bibnamefont {Yu}},\ }\href
  {\doibase 10.1007/JHEP05(2016)046} {\bibfield  {journal} {\bibinfo  {journal}
  {JHEP}\ }\textbf {\bibinfo {volume} {05}},\ \bibinfo {pages} {046} (\bibinfo
  {year} {2016})},\ \Eprint {http://arxiv.org/abs/1509.01598} {arXiv:1509.01598
  [hep-ph]} \BibitemShut {NoStop}%
\bibitem [{\citenamefont {Hochberg}\ \emph
  {et~al.}(2016{\natexlab{a}})\citenamefont {Hochberg}, \citenamefont {Zhao},\
  and\ \citenamefont {Zurek}}]{Hochberg:2015pha}%
  \BibitemOpen
  \bibfield  {author} {\bibinfo {author} {\bibfnamefont {Y.}~\bibnamefont
  {Hochberg}}, \bibinfo {author} {\bibfnamefont {Y.}~\bibnamefont {Zhao}}, \
  and\ \bibinfo {author} {\bibfnamefont {K.~M.}\ \bibnamefont {Zurek}},\ }\href
  {\doibase 10.1103/PhysRevLett.116.011301} {\bibfield  {journal} {\bibinfo
  {journal} {Phys. Rev. Lett.}\ }\textbf {\bibinfo {volume} {116}},\ \bibinfo
  {pages} {011301} (\bibinfo {year} {2016}{\natexlab{a}})},\ \Eprint
  {http://arxiv.org/abs/1504.07237} {arXiv:1504.07237 [hep-ph]} \BibitemShut
  {NoStop}%
\bibitem [{\citenamefont {Derenzo}\ \emph {et~al.}(2017)\citenamefont
  {Derenzo}, \citenamefont {Essig}, \citenamefont {Massari}, \citenamefont
  {Soto},\ and\ \citenamefont {Yu}}]{Derenzo:2016fse}%
  \BibitemOpen
  \bibfield  {author} {\bibinfo {author} {\bibfnamefont {S.}~\bibnamefont
  {Derenzo}}, \bibinfo {author} {\bibfnamefont {R.}~\bibnamefont {Essig}},
  \bibinfo {author} {\bibfnamefont {A.}~\bibnamefont {Massari}}, \bibinfo
  {author} {\bibfnamefont {A.}~\bibnamefont {Soto}}, \ and\ \bibinfo {author}
  {\bibfnamefont {T.-T.}\ \bibnamefont {Yu}},\ }\href {\doibase
  10.1103/PhysRevD.96.016026} {\bibfield  {journal} {\bibinfo  {journal} {Phys.
  Rev. D}\ }\textbf {\bibinfo {volume} {96}},\ \bibinfo {pages} {016026}
  (\bibinfo {year} {2017})},\ \Eprint {http://arxiv.org/abs/1607.01009}
  {arXiv:1607.01009 [hep-ph]} \BibitemShut {NoStop}%
\bibitem [{\citenamefont {Bloch}\ \emph {et~al.}(2017)\citenamefont {Bloch},
  \citenamefont {Essig}, \citenamefont {Tobioka}, \citenamefont {Volansky},\
  and\ \citenamefont {Yu}}]{Bloch:2016sjj}%
  \BibitemOpen
  \bibfield  {author} {\bibinfo {author} {\bibfnamefont {I.~M.}\ \bibnamefont
  {Bloch}}, \bibinfo {author} {\bibfnamefont {R.}~\bibnamefont {Essig}},
  \bibinfo {author} {\bibfnamefont {K.}~\bibnamefont {Tobioka}}, \bibinfo
  {author} {\bibfnamefont {T.}~\bibnamefont {Volansky}}, \ and\ \bibinfo
  {author} {\bibfnamefont {T.-T.}\ \bibnamefont {Yu}},\ }\href {\doibase
  10.1007/JHEP06(2017)087} {\bibfield  {journal} {\bibinfo  {journal} {JHEP}\
  }\textbf {\bibinfo {volume} {06}},\ \bibinfo {pages} {087} (\bibinfo {year}
  {2017})},\ \Eprint {http://arxiv.org/abs/1608.02123} {arXiv:1608.02123
  [hep-ph]} \BibitemShut {NoStop}%
\bibitem [{\citenamefont {Hochberg}\ \emph {et~al.}(2017)\citenamefont
  {Hochberg}, \citenamefont {Kahn}, \citenamefont {Lisanti}, \citenamefont
  {Tully},\ and\ \citenamefont {Zurek}}]{Hochberg:2016ntt}%
  \BibitemOpen
  \bibfield  {author} {\bibinfo {author} {\bibfnamefont {Y.}~\bibnamefont
  {Hochberg}}, \bibinfo {author} {\bibfnamefont {Y.}~\bibnamefont {Kahn}},
  \bibinfo {author} {\bibfnamefont {M.}~\bibnamefont {Lisanti}}, \bibinfo
  {author} {\bibfnamefont {C.~G.}\ \bibnamefont {Tully}}, \ and\ \bibinfo
  {author} {\bibfnamefont {K.~M.}\ \bibnamefont {Zurek}},\ }\href {\doibase
  10.1016/j.physletb.2017.06.051} {\bibfield  {journal} {\bibinfo  {journal}
  {Phys. Lett. B}\ }\textbf {\bibinfo {volume} {772}},\ \bibinfo {pages} {239}
  (\bibinfo {year} {2017})},\ \Eprint {http://arxiv.org/abs/1606.08849}
  {arXiv:1606.08849 [hep-ph]} \BibitemShut {NoStop}%
\bibitem [{\citenamefont {Hochberg}\ \emph
  {et~al.}(2016{\natexlab{b}})\citenamefont {Hochberg}, \citenamefont {Lin},\
  and\ \citenamefont {Zurek}}]{Hochberg:2016ajh}%
  \BibitemOpen
  \bibfield  {author} {\bibinfo {author} {\bibfnamefont {Y.}~\bibnamefont
  {Hochberg}}, \bibinfo {author} {\bibfnamefont {T.}~\bibnamefont {Lin}}, \
  and\ \bibinfo {author} {\bibfnamefont {K.~M.}\ \bibnamefont {Zurek}},\ }\href
  {\doibase 10.1103/PhysRevD.94.015019} {\bibfield  {journal} {\bibinfo
  {journal} {Phys. Rev. D}\ }\textbf {\bibinfo {volume} {94}},\ \bibinfo
  {pages} {015019} (\bibinfo {year} {2016}{\natexlab{b}})},\ \Eprint
  {http://arxiv.org/abs/1604.06800} {arXiv:1604.06800 [hep-ph]} \BibitemShut
  {NoStop}%
\bibitem [{\citenamefont {Kouvaris}\ and\ \citenamefont
  {Pradler}(2017)}]{Kouvaris:2016afs}%
  \BibitemOpen
  \bibfield  {author} {\bibinfo {author} {\bibfnamefont {C.}~\bibnamefont
  {Kouvaris}}\ and\ \bibinfo {author} {\bibfnamefont {J.}~\bibnamefont
  {Pradler}},\ }\href {\doibase 10.1103/PhysRevLett.118.031803} {\bibfield
  {journal} {\bibinfo  {journal} {Phys. Rev. Lett.}\ }\textbf {\bibinfo
  {volume} {118}},\ \bibinfo {pages} {031803} (\bibinfo {year} {2017})},\
  \Eprint {http://arxiv.org/abs/1607.01789} {arXiv:1607.01789 [hep-ph]}
  \BibitemShut {NoStop}%
\bibitem [{\citenamefont {Essig}\ \emph {et~al.}(2017)\citenamefont {Essig},
  \citenamefont {Volansky},\ and\ \citenamefont {Yu}}]{Essig:2017kqs}%
  \BibitemOpen
  \bibfield  {author} {\bibinfo {author} {\bibfnamefont {R.}~\bibnamefont
  {Essig}}, \bibinfo {author} {\bibfnamefont {T.}~\bibnamefont {Volansky}}, \
  and\ \bibinfo {author} {\bibfnamefont {T.-T.}\ \bibnamefont {Yu}},\ }\href
  {\doibase 10.1103/PhysRevD.96.043017} {\bibfield  {journal} {\bibinfo
  {journal} {Phys. Rev. D}\ }\textbf {\bibinfo {volume} {96}},\ \bibinfo
  {pages} {043017} (\bibinfo {year} {2017})},\ \Eprint
  {http://arxiv.org/abs/1703.00910} {arXiv:1703.00910 [hep-ph]} \BibitemShut
  {NoStop}%
\bibitem [{\citenamefont {Budnik}\ \emph {et~al.}(2018)\citenamefont {Budnik},
  \citenamefont {Chesnovsky}, \citenamefont {Slone},\ and\ \citenamefont
  {Volansky}}]{Budnik:2017sbu}%
  \BibitemOpen
  \bibfield  {author} {\bibinfo {author} {\bibfnamefont {R.}~\bibnamefont
  {Budnik}}, \bibinfo {author} {\bibfnamefont {O.}~\bibnamefont {Chesnovsky}},
  \bibinfo {author} {\bibfnamefont {O.}~\bibnamefont {Slone}}, \ and\ \bibinfo
  {author} {\bibfnamefont {T.}~\bibnamefont {Volansky}},\ }\href {\doibase
  10.1016/j.physletb.2018.04.063} {\bibfield  {journal} {\bibinfo  {journal}
  {Phys. Lett. B}\ }\textbf {\bibinfo {volume} {782}},\ \bibinfo {pages} {242}
  (\bibinfo {year} {2018})},\ \Eprint {http://arxiv.org/abs/1705.03016}
  {arXiv:1705.03016 [hep-ph]} \BibitemShut {NoStop}%
\bibitem [{\citenamefont {Bunting}\ \emph {et~al.}(2017)\citenamefont
  {Bunting}, \citenamefont {Gratta}, \citenamefont {Melia},\ and\ \citenamefont
  {Rajendran}}]{Bunting:2017net}%
  \BibitemOpen
  \bibfield  {author} {\bibinfo {author} {\bibfnamefont {P.~C.}\ \bibnamefont
  {Bunting}}, \bibinfo {author} {\bibfnamefont {G.}~\bibnamefont {Gratta}},
  \bibinfo {author} {\bibfnamefont {T.}~\bibnamefont {Melia}}, \ and\ \bibinfo
  {author} {\bibfnamefont {S.}~\bibnamefont {Rajendran}},\ }\href {\doibase
  10.1103/PhysRevD.95.095001} {\bibfield  {journal} {\bibinfo  {journal} {Phys.
  Rev. D}\ }\textbf {\bibinfo {volume} {95}},\ \bibinfo {pages} {095001}
  (\bibinfo {year} {2017})},\ \Eprint {http://arxiv.org/abs/1701.06566}
  {arXiv:1701.06566 [hep-ph]} \BibitemShut {NoStop}%
\bibitem [{\citenamefont {Knapen}\ \emph {et~al.}(2018)\citenamefont {Knapen},
  \citenamefont {Lin}, \citenamefont {Pyle},\ and\ \citenamefont
  {Zurek}}]{Knapen:2017ekk}%
  \BibitemOpen
  \bibfield  {author} {\bibinfo {author} {\bibfnamefont {S.}~\bibnamefont
  {Knapen}}, \bibinfo {author} {\bibfnamefont {T.}~\bibnamefont {Lin}},
  \bibinfo {author} {\bibfnamefont {M.}~\bibnamefont {Pyle}}, \ and\ \bibinfo
  {author} {\bibfnamefont {K.~M.}\ \bibnamefont {Zurek}},\ }\href {\doibase
  10.1016/j.physletb.2018.08.064} {\bibfield  {journal} {\bibinfo  {journal}
  {Phys. Lett. B}\ }\textbf {\bibinfo {volume} {785}},\ \bibinfo {pages} {386}
  (\bibinfo {year} {2018})},\ \Eprint {http://arxiv.org/abs/1712.06598}
  {arXiv:1712.06598 [hep-ph]} \BibitemShut {NoStop}%
\bibitem [{\citenamefont {Hochberg}\ \emph {et~al.}(2018)\citenamefont
  {Hochberg}, \citenamefont {Kahn}, \citenamefont {Lisanti}, \citenamefont
  {Zurek}, \citenamefont {Grushin}, \citenamefont {Ilan}, \citenamefont
  {Griffin}, \citenamefont {Liu}, \citenamefont {Weber},\ and\ \citenamefont
  {Neaton}}]{Hochberg:2017wce}%
  \BibitemOpen
  \bibfield  {author} {\bibinfo {author} {\bibfnamefont {Y.}~\bibnamefont
  {Hochberg}}, \bibinfo {author} {\bibfnamefont {Y.}~\bibnamefont {Kahn}},
  \bibinfo {author} {\bibfnamefont {M.}~\bibnamefont {Lisanti}}, \bibinfo
  {author} {\bibfnamefont {K.~M.}\ \bibnamefont {Zurek}}, \bibinfo {author}
  {\bibfnamefont {A.~G.}\ \bibnamefont {Grushin}}, \bibinfo {author}
  {\bibfnamefont {R.}~\bibnamefont {Ilan}}, \bibinfo {author} {\bibfnamefont
  {S.~M.}\ \bibnamefont {Griffin}}, \bibinfo {author} {\bibfnamefont {Z.-F.}\
  \bibnamefont {Liu}}, \bibinfo {author} {\bibfnamefont {S.~F.}\ \bibnamefont
  {Weber}}, \ and\ \bibinfo {author} {\bibfnamefont {J.~B.}\ \bibnamefont
  {Neaton}},\ }\href {\doibase 10.1103/PhysRevD.97.015004} {\bibfield
  {journal} {\bibinfo  {journal} {Phys. Rev. D}\ }\textbf {\bibinfo {volume}
  {97}},\ \bibinfo {pages} {015004} (\bibinfo {year} {2018})},\ \Eprint
  {http://arxiv.org/abs/1708.08929} {arXiv:1708.08929 [hep-ph]} \BibitemShut
  {NoStop}%
\bibitem [{\citenamefont {Hertel}\ \emph {et~al.}(2019)\citenamefont {Hertel},
  \citenamefont {Biekert}, \citenamefont {Lin}, \citenamefont {Velan},\ and\
  \citenamefont {McKinsey}}]{Hertel:2018aal}%
  \BibitemOpen
  \bibfield  {author} {\bibinfo {author} {\bibfnamefont {S.~A.}\ \bibnamefont
  {Hertel}}, \bibinfo {author} {\bibfnamefont {A.}~\bibnamefont {Biekert}},
  \bibinfo {author} {\bibfnamefont {J.}~\bibnamefont {Lin}}, \bibinfo {author}
  {\bibfnamefont {V.}~\bibnamefont {Velan}}, \ and\ \bibinfo {author}
  {\bibfnamefont {D.~N.}\ \bibnamefont {McKinsey}},\ }\href {\doibase
  10.1103/PhysRevD.100.092007} {\bibfield  {journal} {\bibinfo  {journal}
  {Phys. Rev. D}\ }\textbf {\bibinfo {volume} {100}},\ \bibinfo {pages}
  {092007} (\bibinfo {year} {2019})},\ \Eprint
  {http://arxiv.org/abs/1810.06283} {arXiv:1810.06283 [physics.ins-det]}
  \BibitemShut {NoStop}%
\bibitem [{\citenamefont {Dolan}\ \emph {et~al.}(2018)\citenamefont {Dolan},
  \citenamefont {Kahlhoefer},\ and\ \citenamefont {McCabe}}]{Dolan:2017xbu}%
  \BibitemOpen
  \bibfield  {author} {\bibinfo {author} {\bibfnamefont {M.~J.}\ \bibnamefont
  {Dolan}}, \bibinfo {author} {\bibfnamefont {F.}~\bibnamefont {Kahlhoefer}}, \
  and\ \bibinfo {author} {\bibfnamefont {C.}~\bibnamefont {McCabe}},\ }\href
  {\doibase 10.1103/PhysRevLett.121.101801} {\bibfield  {journal} {\bibinfo
  {journal} {Phys. Rev. Lett.}\ }\textbf {\bibinfo {volume} {121}},\ \bibinfo
  {pages} {101801} (\bibinfo {year} {2018})},\ \Eprint
  {http://arxiv.org/abs/1711.09906} {arXiv:1711.09906 [hep-ph]} \BibitemShut
  {NoStop}%
\bibitem [{\citenamefont {Bringmann}\ and\ \citenamefont
  {Pospelov}(2019)}]{Bringmann:2018cvk}%
  \BibitemOpen
  \bibfield  {author} {\bibinfo {author} {\bibfnamefont {T.}~\bibnamefont
  {Bringmann}}\ and\ \bibinfo {author} {\bibfnamefont {M.}~\bibnamefont
  {Pospelov}},\ }\href {\doibase 10.1103/PhysRevLett.122.171801} {\bibfield
  {journal} {\bibinfo  {journal} {Phys. Rev. Lett.}\ }\textbf {\bibinfo
  {volume} {122}},\ \bibinfo {pages} {171801} (\bibinfo {year} {2019})},\
  \Eprint {http://arxiv.org/abs/1810.10543} {arXiv:1810.10543 [hep-ph]}
  \BibitemShut {NoStop}%
\bibitem [{\citenamefont {Emken}\ \emph {et~al.}(2019)\citenamefont {Emken},
  \citenamefont {Essig}, \citenamefont {Kouvaris},\ and\ \citenamefont
  {Sholapurkar}}]{Emken:2019tni}%
  \BibitemOpen
  \bibfield  {author} {\bibinfo {author} {\bibfnamefont {T.}~\bibnamefont
  {Emken}}, \bibinfo {author} {\bibfnamefont {R.}~\bibnamefont {Essig}},
  \bibinfo {author} {\bibfnamefont {C.}~\bibnamefont {Kouvaris}}, \ and\
  \bibinfo {author} {\bibfnamefont {M.}~\bibnamefont {Sholapurkar}},\ }\href
  {\doibase 10.1088/1475-7516/2019/09/070} {\bibfield  {journal} {\bibinfo
  {journal} {JCAP}\ }\textbf {\bibinfo {volume} {09}},\ \bibinfo {pages} {070}
  (\bibinfo {year} {2019})},\ \Eprint {http://arxiv.org/abs/1905.06348}
  {arXiv:1905.06348 [hep-ph]} \BibitemShut {NoStop}%
\bibitem [{\citenamefont {Essig}\ \emph {et~al.}(2019)\citenamefont {Essig},
  \citenamefont {P\'erez-R\'\i{}os}, \citenamefont {Ramani},\ and\
  \citenamefont {Slone}}]{Essig:2019kfe}%
  \BibitemOpen
  \bibfield  {author} {\bibinfo {author} {\bibfnamefont {R.}~\bibnamefont
  {Essig}}, \bibinfo {author} {\bibfnamefont {J.}~\bibnamefont
  {P\'erez-R\'\i{}os}}, \bibinfo {author} {\bibfnamefont {H.}~\bibnamefont
  {Ramani}}, \ and\ \bibinfo {author} {\bibfnamefont {O.}~\bibnamefont
  {Slone}},\ }\href {\doibase 10.1103/PhysRevResearch.1.033105} {\bibfield
  {journal} {\bibinfo  {journal} {Phys. Rev. Research.}\ }\textbf {\bibinfo
  {volume} {1}},\ \bibinfo {pages} {033105} (\bibinfo {year} {2019})},\ \Eprint
  {http://arxiv.org/abs/1907.07682} {arXiv:1907.07682 [hep-ph]} \BibitemShut
  {NoStop}%
\bibitem [{\citenamefont {Ema}\ \emph {et~al.}(2019)\citenamefont {Ema},
  \citenamefont {Sala},\ and\ \citenamefont {Sato}}]{Ema:2018bih}%
  \BibitemOpen
  \bibfield  {author} {\bibinfo {author} {\bibfnamefont {Y.}~\bibnamefont
  {Ema}}, \bibinfo {author} {\bibfnamefont {F.}~\bibnamefont {Sala}}, \ and\
  \bibinfo {author} {\bibfnamefont {R.}~\bibnamefont {Sato}},\ }\href {\doibase
  10.1103/PhysRevLett.122.181802} {\bibfield  {journal} {\bibinfo  {journal}
  {Phys. Rev. Lett.}\ }\textbf {\bibinfo {volume} {122}},\ \bibinfo {pages}
  {181802} (\bibinfo {year} {2019})},\ \Eprint
  {http://arxiv.org/abs/1811.00520} {arXiv:1811.00520 [hep-ph]} \BibitemShut
  {NoStop}%
\bibitem [{\citenamefont {Bell}\ \emph
  {et~al.}(2020{\natexlab{a}})\citenamefont {Bell}, \citenamefont {Dent},
  \citenamefont {Newstead}, \citenamefont {Sabharwal},\ and\ \citenamefont
  {Weiler}}]{Bell:2019egg}%
  \BibitemOpen
  \bibfield  {author} {\bibinfo {author} {\bibfnamefont {N.~F.}\ \bibnamefont
  {Bell}}, \bibinfo {author} {\bibfnamefont {J.~B.}\ \bibnamefont {Dent}},
  \bibinfo {author} {\bibfnamefont {J.~L.}\ \bibnamefont {Newstead}}, \bibinfo
  {author} {\bibfnamefont {S.}~\bibnamefont {Sabharwal}}, \ and\ \bibinfo
  {author} {\bibfnamefont {T.~J.}\ \bibnamefont {Weiler}},\ }\href {\doibase
  10.1103/PhysRevD.101.015012} {\bibfield  {journal} {\bibinfo  {journal}
  {Phys. Rev. D}\ }\textbf {\bibinfo {volume} {101}},\ \bibinfo {pages}
  {015012} (\bibinfo {year} {2020}{\natexlab{a}})},\ \Eprint
  {http://arxiv.org/abs/1905.00046} {arXiv:1905.00046 [hep-ph]} \BibitemShut
  {NoStop}%
\bibitem [{\citenamefont {Trickle}\ \emph
  {et~al.}(2020{\natexlab{a}})\citenamefont {Trickle}, \citenamefont {Zhang},\
  and\ \citenamefont {Zurek}}]{Trickle:2019ovy}%
  \BibitemOpen
  \bibfield  {author} {\bibinfo {author} {\bibfnamefont {T.}~\bibnamefont
  {Trickle}}, \bibinfo {author} {\bibfnamefont {Z.}~\bibnamefont {Zhang}}, \
  and\ \bibinfo {author} {\bibfnamefont {K.~M.}\ \bibnamefont {Zurek}},\ }\href
  {\doibase 10.1103/PhysRevLett.124.201801} {\bibfield  {journal} {\bibinfo
  {journal} {Phys. Rev. Lett.}\ }\textbf {\bibinfo {volume} {124}},\ \bibinfo
  {pages} {201801} (\bibinfo {year} {2020}{\natexlab{a}})},\ \Eprint
  {http://arxiv.org/abs/1905.13744} {arXiv:1905.13744 [hep-ph]} \BibitemShut
  {NoStop}%
\bibitem [{\citenamefont {Trickle}\ \emph
  {et~al.}(2020{\natexlab{b}})\citenamefont {Trickle}, \citenamefont {Zhang},
  \citenamefont {Zurek}, \citenamefont {Inzani},\ and\ \citenamefont
  {Griffin}}]{Trickle:2019nya}%
  \BibitemOpen
  \bibfield  {author} {\bibinfo {author} {\bibfnamefont {T.}~\bibnamefont
  {Trickle}}, \bibinfo {author} {\bibfnamefont {Z.}~\bibnamefont {Zhang}},
  \bibinfo {author} {\bibfnamefont {K.~M.}\ \bibnamefont {Zurek}}, \bibinfo
  {author} {\bibfnamefont {K.}~\bibnamefont {Inzani}}, \ and\ \bibinfo {author}
  {\bibfnamefont {S.}~\bibnamefont {Griffin}},\ }\href {\doibase
  10.1007/JHEP03(2020)036} {\bibfield  {journal} {\bibinfo  {journal} {JHEP}\
  }\textbf {\bibinfo {volume} {03}},\ \bibinfo {pages} {036} (\bibinfo {year}
  {2020}{\natexlab{b}})},\ \Eprint {http://arxiv.org/abs/1910.08092}
  {arXiv:1910.08092 [hep-ph]} \BibitemShut {NoStop}%
\bibitem [{\citenamefont {Griffin}\ \emph {et~al.}(2020)\citenamefont
  {Griffin}, \citenamefont {Inzani}, \citenamefont {Trickle}, \citenamefont
  {Zhang},\ and\ \citenamefont {Zurek}}]{Griffin:2019mvc}%
  \BibitemOpen
  \bibfield  {author} {\bibinfo {author} {\bibfnamefont {S.~M.}\ \bibnamefont
  {Griffin}}, \bibinfo {author} {\bibfnamefont {K.}~\bibnamefont {Inzani}},
  \bibinfo {author} {\bibfnamefont {T.}~\bibnamefont {Trickle}}, \bibinfo
  {author} {\bibfnamefont {Z.}~\bibnamefont {Zhang}}, \ and\ \bibinfo {author}
  {\bibfnamefont {K.~M.}\ \bibnamefont {Zurek}},\ }\href {\doibase
  10.1103/PhysRevD.101.055004} {\bibfield  {journal} {\bibinfo  {journal}
  {Phys. Rev. D}\ }\textbf {\bibinfo {volume} {101}},\ \bibinfo {pages}
  {055004} (\bibinfo {year} {2020})},\ \Eprint
  {http://arxiv.org/abs/1910.10716} {arXiv:1910.10716 [hep-ph]} \BibitemShut
  {NoStop}%
\bibitem [{\citenamefont {Baxter}\ \emph {et~al.}(2020)\citenamefont {Baxter},
  \citenamefont {Kahn},\ and\ \citenamefont {Krnjaic}}]{Baxter:2019pnz}%
  \BibitemOpen
  \bibfield  {author} {\bibinfo {author} {\bibfnamefont {D.}~\bibnamefont
  {Baxter}}, \bibinfo {author} {\bibfnamefont {Y.}~\bibnamefont {Kahn}}, \ and\
  \bibinfo {author} {\bibfnamefont {G.}~\bibnamefont {Krnjaic}},\ }\href
  {\doibase 10.1103/PhysRevD.101.076014} {\bibfield  {journal} {\bibinfo
  {journal} {Phys. Rev. D}\ }\textbf {\bibinfo {volume} {101}},\ \bibinfo
  {pages} {076014} (\bibinfo {year} {2020})},\ \Eprint
  {http://arxiv.org/abs/1908.00012} {arXiv:1908.00012 [hep-ph]} \BibitemShut
  {NoStop}%
\bibitem [{\citenamefont {Kurinsky}\ \emph {et~al.}(2019)\citenamefont
  {Kurinsky}, \citenamefont {Yu}, \citenamefont {Hochberg},\ and\ \citenamefont
  {Cabrera}}]{Kurinsky:2019pgb}%
  \BibitemOpen
  \bibfield  {author} {\bibinfo {author} {\bibfnamefont {N.~A.}\ \bibnamefont
  {Kurinsky}}, \bibinfo {author} {\bibfnamefont {T.~C.}\ \bibnamefont {Yu}},
  \bibinfo {author} {\bibfnamefont {Y.}~\bibnamefont {Hochberg}}, \ and\
  \bibinfo {author} {\bibfnamefont {B.}~\bibnamefont {Cabrera}},\ }\href
  {\doibase 10.1103/PhysRevD.99.123005} {\bibfield  {journal} {\bibinfo
  {journal} {Phys. Rev. D}\ }\textbf {\bibinfo {volume} {99}},\ \bibinfo
  {pages} {123005} (\bibinfo {year} {2019})},\ \Eprint
  {http://arxiv.org/abs/1901.07569} {arXiv:1901.07569 [hep-ex]} \BibitemShut
  {NoStop}%
\bibitem [{\citenamefont {Catena}\ \emph {et~al.}(2020)\citenamefont {Catena},
  \citenamefont {Emken}, \citenamefont {Spaldin},\ and\ \citenamefont
  {Tarantino}}]{Catena:2019gfa}%
  \BibitemOpen
  \bibfield  {author} {\bibinfo {author} {\bibfnamefont {R.}~\bibnamefont
  {Catena}}, \bibinfo {author} {\bibfnamefont {T.}~\bibnamefont {Emken}},
  \bibinfo {author} {\bibfnamefont {N.~A.}\ \bibnamefont {Spaldin}}, \ and\
  \bibinfo {author} {\bibfnamefont {W.}~\bibnamefont {Tarantino}},\ }\href
  {\doibase 10.1103/PhysRevResearch.2.033195} {\bibfield  {journal} {\bibinfo
  {journal} {Phys. Rev. Res.}\ }\textbf {\bibinfo {volume} {2}},\ \bibinfo
  {pages} {033195} (\bibinfo {year} {2020})},\ \Eprint
  {http://arxiv.org/abs/1912.08204} {arXiv:1912.08204 [hep-ph]} \BibitemShut
  {NoStop}%
\bibitem [{\citenamefont {Griffin}\ \emph {et~al.}(2021)\citenamefont
  {Griffin}, \citenamefont {Hochberg}, \citenamefont {Inzani}, \citenamefont
  {Kurinsky}, \citenamefont {Lin},\ and\ \citenamefont
  {Chin}}]{Griffin:2020lgd}%
  \BibitemOpen
  \bibfield  {author} {\bibinfo {author} {\bibfnamefont {S.~M.}\ \bibnamefont
  {Griffin}}, \bibinfo {author} {\bibfnamefont {Y.}~\bibnamefont {Hochberg}},
  \bibinfo {author} {\bibfnamefont {K.}~\bibnamefont {Inzani}}, \bibinfo
  {author} {\bibfnamefont {N.}~\bibnamefont {Kurinsky}}, \bibinfo {author}
  {\bibfnamefont {T.}~\bibnamefont {Lin}}, \ and\ \bibinfo {author}
  {\bibfnamefont {T.}~\bibnamefont {Chin}},\ }\href {\doibase
  10.1103/PhysRevD.103.075002} {\bibfield  {journal} {\bibinfo  {journal}
  {Phys. Rev. D}\ }\textbf {\bibinfo {volume} {103}},\ \bibinfo {pages}
  {075002} (\bibinfo {year} {2021})},\ \Eprint
  {http://arxiv.org/abs/2008.08560} {arXiv:2008.08560 [hep-ph]} \BibitemShut
  {NoStop}%
\bibitem [{\citenamefont {Flambaum}\ \emph {et~al.}(2020)\citenamefont
  {Flambaum}, \citenamefont {Su}, \citenamefont {Wu},\ and\ \citenamefont
  {Zhu}}]{Flambaum:2020xxo}%
  \BibitemOpen
  \bibfield  {author} {\bibinfo {author} {\bibfnamefont {V.~V.}\ \bibnamefont
  {Flambaum}}, \bibinfo {author} {\bibfnamefont {L.}~\bibnamefont {Su}},
  \bibinfo {author} {\bibfnamefont {L.}~\bibnamefont {Wu}}, \ and\ \bibinfo
  {author} {\bibfnamefont {B.}~\bibnamefont {Zhu}},\ }\href@noop {} {\
  (\bibinfo {year} {2020})},\ \Eprint {http://arxiv.org/abs/2012.09751}
  {arXiv:2012.09751 [hep-ph]} \BibitemShut {NoStop}%
\bibitem [{\citenamefont {Bell}\ \emph {et~al.}(2021)\citenamefont {Bell},
  \citenamefont {Dent}, \citenamefont {Dutta}, \citenamefont {Ghosh},
  \citenamefont {Kumar},\ and\ \citenamefont {Newstead}}]{Bell:2021zkr}%
  \BibitemOpen
  \bibfield  {author} {\bibinfo {author} {\bibfnamefont {N.~F.}\ \bibnamefont
  {Bell}}, \bibinfo {author} {\bibfnamefont {J.~B.}\ \bibnamefont {Dent}},
  \bibinfo {author} {\bibfnamefont {B.}~\bibnamefont {Dutta}}, \bibinfo
  {author} {\bibfnamefont {S.}~\bibnamefont {Ghosh}}, \bibinfo {author}
  {\bibfnamefont {J.}~\bibnamefont {Kumar}}, \ and\ \bibinfo {author}
  {\bibfnamefont {J.~L.}\ \bibnamefont {Newstead}},\ }\href@noop {} {\
  (\bibinfo {year} {2021})},\ \Eprint {http://arxiv.org/abs/2103.05890}
  {arXiv:2103.05890 [hep-ph]} \BibitemShut {NoStop}%
\bibitem [{\citenamefont {Migdal}(1941)}]{Migdal:1941}%
  \BibitemOpen
  \bibfield  {author} {\bibinfo {author} {\bibfnamefont {A.}~\bibnamefont
  {Migdal}},\ }\href@noop {} {\bibfield  {journal} {\bibinfo  {journal}
  {J.Phys.(USSR)}\ }\textbf {\bibinfo {volume} {4}},\ \bibinfo {pages} {449}
  (\bibinfo {year} {1941})}\BibitemShut {NoStop}%
\bibitem [{\citenamefont {Vergados}\ and\ \citenamefont
  {Ejiri}(2005)}]{Vergados:2004bm}%
  \BibitemOpen
  \bibfield  {author} {\bibinfo {author} {\bibfnamefont {J.~D.}\ \bibnamefont
  {Vergados}}\ and\ \bibinfo {author} {\bibfnamefont {H.}~\bibnamefont
  {Ejiri}},\ }\href {\doibase 10.1016/j.physletb.2004.11.085} {\bibfield
  {journal} {\bibinfo  {journal} {Phys. Lett. B}\ }\textbf {\bibinfo {volume}
  {606}},\ \bibinfo {pages} {313} (\bibinfo {year} {2005})},\ \Eprint
  {http://arxiv.org/abs/hep-ph/0401151} {arXiv:hep-ph/0401151} \BibitemShut
  {NoStop}%
\bibitem [{\citenamefont {Bernabei}\ \emph {et~al.}(2007)\citenamefont
  {Bernabei} \emph {et~al.}}]{Bernabei:2007jz}%
  \BibitemOpen
  \bibfield  {author} {\bibinfo {author} {\bibfnamefont {R.}~\bibnamefont
  {Bernabei}} \emph {et~al.},\ }\href {\doibase 10.1142/S0217751X07037093}
  {\bibfield  {journal} {\bibinfo  {journal} {Int. J. Mod. Phys. A}\ }\textbf
  {\bibinfo {volume} {22}},\ \bibinfo {pages} {3155} (\bibinfo {year}
  {2007})},\ \Eprint {http://arxiv.org/abs/0706.1421} {arXiv:0706.1421
  [astro-ph]} \BibitemShut {NoStop}%
\bibitem [{\citenamefont {Ibe}\ \emph {et~al.}(2018)\citenamefont {Ibe},
  \citenamefont {Nakano}, \citenamefont {Shoji},\ and\ \citenamefont
  {Suzuki}}]{Ibe:2017yqa}%
  \BibitemOpen
  \bibfield  {author} {\bibinfo {author} {\bibfnamefont {M.}~\bibnamefont
  {Ibe}}, \bibinfo {author} {\bibfnamefont {W.}~\bibnamefont {Nakano}},
  \bibinfo {author} {\bibfnamefont {Y.}~\bibnamefont {Shoji}}, \ and\ \bibinfo
  {author} {\bibfnamefont {K.}~\bibnamefont {Suzuki}},\ }\href {\doibase
  10.1007/JHEP03(2018)194} {\bibfield  {journal} {\bibinfo  {journal} {JHEP}\
  }\textbf {\bibinfo {volume} {03}},\ \bibinfo {pages} {194} (\bibinfo {year}
  {2018})},\ \Eprint {http://arxiv.org/abs/1707.07258} {arXiv:1707.07258
  [hep-ph]} \BibitemShut {NoStop}%
\bibitem [{\citenamefont {Essig}\ \emph {et~al.}(2020)\citenamefont {Essig},
  \citenamefont {Pradler}, \citenamefont {Sholapurkar},\ and\ \citenamefont
  {Yu}}]{Essig:2019xkx}%
  \BibitemOpen
  \bibfield  {author} {\bibinfo {author} {\bibfnamefont {R.}~\bibnamefont
  {Essig}}, \bibinfo {author} {\bibfnamefont {J.}~\bibnamefont {Pradler}},
  \bibinfo {author} {\bibfnamefont {M.}~\bibnamefont {Sholapurkar}}, \ and\
  \bibinfo {author} {\bibfnamefont {T.-T.}\ \bibnamefont {Yu}},\ }\href
  {\doibase 10.1103/PhysRevLett.124.021801} {\bibfield  {journal} {\bibinfo
  {journal} {Phys. Rev. Lett.}\ }\textbf {\bibinfo {volume} {124}},\ \bibinfo
  {pages} {021801} (\bibinfo {year} {2020})},\ \Eprint
  {http://arxiv.org/abs/1908.10881} {arXiv:1908.10881 [hep-ph]} \BibitemShut
  {NoStop}%
\bibitem [{\citenamefont {Liu}\ \emph {et~al.}(2020)\citenamefont {Liu},
  \citenamefont {Wu}, \citenamefont {Chi},\ and\ \citenamefont
  {Chen}}]{Liu:2020pat}%
  \BibitemOpen
  \bibfield  {author} {\bibinfo {author} {\bibfnamefont {C.~P.}\ \bibnamefont
  {Liu}}, \bibinfo {author} {\bibfnamefont {C.-P.}\ \bibnamefont {Wu}},
  \bibinfo {author} {\bibfnamefont {H.-C.}\ \bibnamefont {Chi}}, \ and\
  \bibinfo {author} {\bibfnamefont {J.-W.}\ \bibnamefont {Chen}},\ }\href
  {\doibase 10.1103/PhysRevD.102.121303} {\bibfield  {journal} {\bibinfo
  {journal} {Phys. Rev. D}\ }\textbf {\bibinfo {volume} {102}},\ \bibinfo
  {pages} {121303} (\bibinfo {year} {2020})},\ \Eprint
  {http://arxiv.org/abs/2007.10965} {arXiv:2007.10965 [hep-ph]} \BibitemShut
  {NoStop}%
\bibitem [{\citenamefont {Grilli~di Cortona}\ \emph {et~al.}(2020)\citenamefont
  {Grilli~di Cortona}, \citenamefont {Messina},\ and\ \citenamefont
  {Piacentini}}]{GrillidiCortona:2020owp}%
  \BibitemOpen
  \bibfield  {author} {\bibinfo {author} {\bibfnamefont {G.}~\bibnamefont
  {Grilli~di Cortona}}, \bibinfo {author} {\bibfnamefont {A.}~\bibnamefont
  {Messina}}, \ and\ \bibinfo {author} {\bibfnamefont {S.}~\bibnamefont
  {Piacentini}},\ }\href {\doibase 10.1007/JHEP11(2020)034} {\bibfield
  {journal} {\bibinfo  {journal} {JHEP}\ }\textbf {\bibinfo {volume} {11}},\
  \bibinfo {pages} {034} (\bibinfo {year} {2020})},\ \Eprint
  {http://arxiv.org/abs/2006.02453} {arXiv:2006.02453 [hep-ph]} \BibitemShut
  {NoStop}%
\bibitem [{\citenamefont {Dey}\ \emph {et~al.}(2020)\citenamefont {Dey},
  \citenamefont {Maity},\ and\ \citenamefont {Ray}}]{Dey:2020sai}%
  \BibitemOpen
  \bibfield  {author} {\bibinfo {author} {\bibfnamefont {U.~K.}\ \bibnamefont
  {Dey}}, \bibinfo {author} {\bibfnamefont {T.~N.}\ \bibnamefont {Maity}}, \
  and\ \bibinfo {author} {\bibfnamefont {T.~S.}\ \bibnamefont {Ray}},\ }\href
  {\doibase 10.1016/j.physletb.2020.135900} {\bibfield  {journal} {\bibinfo
  {journal} {Phys. Lett. B}\ }\textbf {\bibinfo {volume} {811}},\ \bibinfo
  {pages} {135900} (\bibinfo {year} {2020})},\ \Eprint
  {http://arxiv.org/abs/2006.12529} {arXiv:2006.12529 [hep-ph]} \BibitemShut
  {NoStop}%
\bibitem [{\citenamefont {Dent}\ \emph {et~al.}(2020)\citenamefont {Dent},
  \citenamefont {Dutta}, \citenamefont {Newstead},\ and\ \citenamefont
  {Shoemaker}}]{Dent:2019krz}%
  \BibitemOpen
  \bibfield  {author} {\bibinfo {author} {\bibfnamefont {J.~B.}\ \bibnamefont
  {Dent}}, \bibinfo {author} {\bibfnamefont {B.}~\bibnamefont {Dutta}},
  \bibinfo {author} {\bibfnamefont {J.~L.}\ \bibnamefont {Newstead}}, \ and\
  \bibinfo {author} {\bibfnamefont {I.~M.}\ \bibnamefont {Shoemaker}},\ }\href
  {\doibase 10.1103/PhysRevD.101.116007} {\bibfield  {journal} {\bibinfo
  {journal} {Phys. Rev. D}\ }\textbf {\bibinfo {volume} {101}},\ \bibinfo
  {pages} {116007} (\bibinfo {year} {2020})},\ \Eprint
  {http://arxiv.org/abs/1907.03782} {arXiv:1907.03782 [hep-ph]} \BibitemShut
  {NoStop}%
\bibitem [{\citenamefont {Cappiello}\ and\ \citenamefont
  {Beacom}(2019)}]{Cappiello:2019qsw}%
  \BibitemOpen
  \bibfield  {author} {\bibinfo {author} {\bibfnamefont {C.}~\bibnamefont
  {Cappiello}}\ and\ \bibinfo {author} {\bibfnamefont {J.~F.}\ \bibnamefont
  {Beacom}},\ }\href {\doibase 10.1103/PhysRevD.100.103011} {\bibfield
  {journal} {\bibinfo  {journal} {Phys. Rev. D}\ }\textbf {\bibinfo {volume}
  {100}},\ \bibinfo {pages} {103011} (\bibinfo {year} {2019})},\ \Eprint
  {http://arxiv.org/abs/1906.11283} {arXiv:1906.11283 [hep-ph]} \BibitemShut
  {NoStop}%
\bibitem [{\citenamefont {Bondarenko}\ \emph {et~al.}(2020)\citenamefont
  {Bondarenko}, \citenamefont {Boyarsky}, \citenamefont {Bringmann},
  \citenamefont {Hufnagel}, \citenamefont {Schmidt-Hoberg},\ and\ \citenamefont
  {Sokolenko}}]{Bondarenko:2019vrb}%
  \BibitemOpen
  \bibfield  {author} {\bibinfo {author} {\bibfnamefont {K.}~\bibnamefont
  {Bondarenko}}, \bibinfo {author} {\bibfnamefont {A.}~\bibnamefont
  {Boyarsky}}, \bibinfo {author} {\bibfnamefont {T.}~\bibnamefont {Bringmann}},
  \bibinfo {author} {\bibfnamefont {M.}~\bibnamefont {Hufnagel}}, \bibinfo
  {author} {\bibfnamefont {K.}~\bibnamefont {Schmidt-Hoberg}}, \ and\ \bibinfo
  {author} {\bibfnamefont {A.}~\bibnamefont {Sokolenko}},\ }\href {\doibase
  10.1007/JHEP03(2020)118} {\bibfield  {journal} {\bibinfo  {journal} {JHEP}\
  }\textbf {\bibinfo {volume} {03}},\ \bibinfo {pages} {118} (\bibinfo {year}
  {2020})},\ \Eprint {http://arxiv.org/abs/1909.08632} {arXiv:1909.08632
  [hep-ph]} \BibitemShut {NoStop}%
\bibitem [{\citenamefont {Guo}\ \emph {et~al.}(2020)\citenamefont {Guo},
  \citenamefont {Tsai}, \citenamefont {Wu},\ and\ \citenamefont
  {Yuan}}]{Guo:2020oum}%
  \BibitemOpen
  \bibfield  {author} {\bibinfo {author} {\bibfnamefont {G.}~\bibnamefont
  {Guo}}, \bibinfo {author} {\bibfnamefont {Y.-L.~S.}\ \bibnamefont {Tsai}},
  \bibinfo {author} {\bibfnamefont {M.-R.}\ \bibnamefont {Wu}}, \ and\ \bibinfo
  {author} {\bibfnamefont {Q.}~\bibnamefont {Yuan}},\ }\href {\doibase
  10.1103/PhysRevD.102.103004} {\bibfield  {journal} {\bibinfo  {journal}
  {Phys. Rev. D}\ }\textbf {\bibinfo {volume} {102}},\ \bibinfo {pages}
  {103004} (\bibinfo {year} {2020})},\ \Eprint
  {http://arxiv.org/abs/2008.12137} {arXiv:2008.12137 [astro-ph.HE]}
  \BibitemShut {NoStop}%
\bibitem [{\citenamefont {Tucker-Smith}\ and\ \citenamefont
  {Weiner}(2001)}]{TuckerSmith:2001hy}%
  \BibitemOpen
  \bibfield  {author} {\bibinfo {author} {\bibfnamefont {D.}~\bibnamefont
  {Tucker-Smith}}\ and\ \bibinfo {author} {\bibfnamefont {N.}~\bibnamefont
  {Weiner}},\ }\href {\doibase 10.1103/PhysRevD.64.043502} {\bibfield
  {journal} {\bibinfo  {journal} {Phys. Rev. D}\ }\textbf {\bibinfo {volume}
  {64}},\ \bibinfo {pages} {043502} (\bibinfo {year} {2001})},\ \Eprint
  {http://arxiv.org/abs/hep-ph/0101138} {arXiv:hep-ph/0101138} \BibitemShut
  {NoStop}%
\bibitem [{\citenamefont {Tucker-Smith}\ and\ \citenamefont
  {Weiner}(2005)}]{TuckerSmith:2004jv}%
  \BibitemOpen
  \bibfield  {author} {\bibinfo {author} {\bibfnamefont {D.}~\bibnamefont
  {Tucker-Smith}}\ and\ \bibinfo {author} {\bibfnamefont {N.}~\bibnamefont
  {Weiner}},\ }\href {\doibase 10.1103/PhysRevD.72.063509} {\bibfield
  {journal} {\bibinfo  {journal} {Phys. Rev. D}\ }\textbf {\bibinfo {volume}
  {72}},\ \bibinfo {pages} {063509} (\bibinfo {year} {2005})},\ \Eprint
  {http://arxiv.org/abs/hep-ph/0402065} {arXiv:hep-ph/0402065} \BibitemShut
  {NoStop}%
\bibitem [{\citenamefont {Finkbeiner}\ and\ \citenamefont
  {Weiner}(2007)}]{Finkbeiner:2007kk}%
  \BibitemOpen
  \bibfield  {author} {\bibinfo {author} {\bibfnamefont {D.~P.}\ \bibnamefont
  {Finkbeiner}}\ and\ \bibinfo {author} {\bibfnamefont {N.}~\bibnamefont
  {Weiner}},\ }\href {\doibase 10.1103/PhysRevD.76.083519} {\bibfield
  {journal} {\bibinfo  {journal} {Phys. Rev. D}\ }\textbf {\bibinfo {volume}
  {76}},\ \bibinfo {pages} {083519} (\bibinfo {year} {2007})},\ \Eprint
  {http://arxiv.org/abs/astro-ph/0702587} {arXiv:astro-ph/0702587} \BibitemShut
  {NoStop}%
\bibitem [{\citenamefont {Arina}\ and\ \citenamefont
  {Fornengo}(2007)}]{Arina:2007tm}%
  \BibitemOpen
  \bibfield  {author} {\bibinfo {author} {\bibfnamefont {C.}~\bibnamefont
  {Arina}}\ and\ \bibinfo {author} {\bibfnamefont {N.}~\bibnamefont
  {Fornengo}},\ }\href {\doibase 10.1088/1126-6708/2007/11/029} {\bibfield
  {journal} {\bibinfo  {journal} {JHEP}\ }\textbf {\bibinfo {volume} {11}},\
  \bibinfo {pages} {029} (\bibinfo {year} {2007})},\ \Eprint
  {http://arxiv.org/abs/0709.4477} {arXiv:0709.4477 [hep-ph]} \BibitemShut
  {NoStop}%
\bibitem [{\citenamefont {Chang}\ \emph {et~al.}(2009)\citenamefont {Chang},
  \citenamefont {Kribs}, \citenamefont {Tucker-Smith},\ and\ \citenamefont
  {Weiner}}]{Chang:2008gd}%
  \BibitemOpen
  \bibfield  {author} {\bibinfo {author} {\bibfnamefont {S.}~\bibnamefont
  {Chang}}, \bibinfo {author} {\bibfnamefont {G.~D.}\ \bibnamefont {Kribs}},
  \bibinfo {author} {\bibfnamefont {D.}~\bibnamefont {Tucker-Smith}}, \ and\
  \bibinfo {author} {\bibfnamefont {N.}~\bibnamefont {Weiner}},\ }\href
  {\doibase 10.1103/PhysRevD.79.043513} {\bibfield  {journal} {\bibinfo
  {journal} {Phys. Rev. D}\ }\textbf {\bibinfo {volume} {79}},\ \bibinfo
  {pages} {043513} (\bibinfo {year} {2009})},\ \Eprint
  {http://arxiv.org/abs/0807.2250} {arXiv:0807.2250 [hep-ph]} \BibitemShut
  {NoStop}%
\bibitem [{\citenamefont {Cui}\ \emph {et~al.}(2009)\citenamefont {Cui},
  \citenamefont {Morrissey}, \citenamefont {Poland},\ and\ \citenamefont
  {Randall}}]{Cui:2009xq}%
  \BibitemOpen
  \bibfield  {author} {\bibinfo {author} {\bibfnamefont {Y.}~\bibnamefont
  {Cui}}, \bibinfo {author} {\bibfnamefont {D.~E.}\ \bibnamefont {Morrissey}},
  \bibinfo {author} {\bibfnamefont {D.}~\bibnamefont {Poland}}, \ and\ \bibinfo
  {author} {\bibfnamefont {L.}~\bibnamefont {Randall}},\ }\href {\doibase
  10.1088/1126-6708/2009/05/076} {\bibfield  {journal} {\bibinfo  {journal}
  {JHEP}\ }\textbf {\bibinfo {volume} {05}},\ \bibinfo {pages} {076} (\bibinfo
  {year} {2009})},\ \Eprint {http://arxiv.org/abs/0901.0557} {arXiv:0901.0557
  [hep-ph]} \BibitemShut {NoStop}%
\bibitem [{\citenamefont {Fox}\ \emph {et~al.}(2011)\citenamefont {Fox},
  \citenamefont {Kribs},\ and\ \citenamefont {Tait}}]{Fox:2010bu}%
  \BibitemOpen
  \bibfield  {author} {\bibinfo {author} {\bibfnamefont {P.~J.}\ \bibnamefont
  {Fox}}, \bibinfo {author} {\bibfnamefont {G.~D.}\ \bibnamefont {Kribs}}, \
  and\ \bibinfo {author} {\bibfnamefont {T.~M.}\ \bibnamefont {Tait}},\ }\href
  {\doibase 10.1103/PhysRevD.83.034007} {\bibfield  {journal} {\bibinfo
  {journal} {Phys. Rev. D}\ }\textbf {\bibinfo {volume} {83}},\ \bibinfo
  {pages} {034007} (\bibinfo {year} {2011})},\ \Eprint
  {http://arxiv.org/abs/1011.1910} {arXiv:1011.1910 [hep-ph]} \BibitemShut
  {NoStop}%
\bibitem [{\citenamefont {Lin}\ and\ \citenamefont
  {Finkbeiner}(2011)}]{Lin:2010sb}%
  \BibitemOpen
  \bibfield  {author} {\bibinfo {author} {\bibfnamefont {T.}~\bibnamefont
  {Lin}}\ and\ \bibinfo {author} {\bibfnamefont {D.~P.}\ \bibnamefont
  {Finkbeiner}},\ }\href {\doibase 10.1103/PhysRevD.83.083510} {\bibfield
  {journal} {\bibinfo  {journal} {Phys. Rev. D}\ }\textbf {\bibinfo {volume}
  {83}},\ \bibinfo {pages} {083510} (\bibinfo {year} {2011})},\ \Eprint
  {http://arxiv.org/abs/1011.3052} {arXiv:1011.3052 [astro-ph.CO]} \BibitemShut
  {NoStop}%
\bibitem [{\citenamefont {De~Simone}\ \emph {et~al.}(2010)\citenamefont
  {De~Simone}, \citenamefont {Sanz},\ and\ \citenamefont
  {Sato}}]{DeSimone:2010tf}%
  \BibitemOpen
  \bibfield  {author} {\bibinfo {author} {\bibfnamefont {A.}~\bibnamefont
  {De~Simone}}, \bibinfo {author} {\bibfnamefont {V.}~\bibnamefont {Sanz}}, \
  and\ \bibinfo {author} {\bibfnamefont {H.~P.}\ \bibnamefont {Sato}},\ }\href
  {\doibase 10.1103/PhysRevLett.105.121802} {\bibfield  {journal} {\bibinfo
  {journal} {Phys. Rev. Lett.}\ }\textbf {\bibinfo {volume} {105}},\ \bibinfo
  {pages} {121802} (\bibinfo {year} {2010})},\ \Eprint
  {http://arxiv.org/abs/1004.1567} {arXiv:1004.1567 [hep-ph]} \BibitemShut
  {NoStop}%
\bibitem [{\citenamefont {An}\ \emph {et~al.}(2012)\citenamefont {An},
  \citenamefont {Dev}, \citenamefont {Cai},\ and\ \citenamefont
  {Mohapatra}}]{An:2011uq}%
  \BibitemOpen
  \bibfield  {author} {\bibinfo {author} {\bibfnamefont {H.}~\bibnamefont
  {An}}, \bibinfo {author} {\bibfnamefont {P.}~\bibnamefont {Dev}}, \bibinfo
  {author} {\bibfnamefont {Y.}~\bibnamefont {Cai}}, \ and\ \bibinfo {author}
  {\bibfnamefont {R.}~\bibnamefont {Mohapatra}},\ }\href {\doibase
  10.1103/PhysRevLett.108.081806} {\bibfield  {journal} {\bibinfo  {journal}
  {Phys. Rev. Lett.}\ }\textbf {\bibinfo {volume} {108}},\ \bibinfo {pages}
  {081806} (\bibinfo {year} {2012})},\ \Eprint {http://arxiv.org/abs/1110.1366}
  {arXiv:1110.1366 [hep-ph]} \BibitemShut {NoStop}%
\bibitem [{\citenamefont {Pospelov}\ \emph {et~al.}(2014)\citenamefont
  {Pospelov}, \citenamefont {Weiner},\ and\ \citenamefont
  {Yavin}}]{Pospelov:2013nea}%
  \BibitemOpen
  \bibfield  {author} {\bibinfo {author} {\bibfnamefont {M.}~\bibnamefont
  {Pospelov}}, \bibinfo {author} {\bibfnamefont {N.}~\bibnamefont {Weiner}}, \
  and\ \bibinfo {author} {\bibfnamefont {I.}~\bibnamefont {Yavin}},\ }\href
  {\doibase 10.1103/PhysRevD.89.055008} {\bibfield  {journal} {\bibinfo
  {journal} {Phys. Rev. D}\ }\textbf {\bibinfo {volume} {89}},\ \bibinfo
  {pages} {055008} (\bibinfo {year} {2014})},\ \Eprint
  {http://arxiv.org/abs/1312.1363} {arXiv:1312.1363 [hep-ph]} \BibitemShut
  {NoStop}%
\bibitem [{\citenamefont {Finkbeiner}\ and\ \citenamefont
  {Weiner}(2016)}]{Finkbeiner:2014sja}%
  \BibitemOpen
  \bibfield  {author} {\bibinfo {author} {\bibfnamefont {D.~P.}\ \bibnamefont
  {Finkbeiner}}\ and\ \bibinfo {author} {\bibfnamefont {N.}~\bibnamefont
  {Weiner}},\ }\href {\doibase 10.1103/PhysRevD.94.083002} {\bibfield
  {journal} {\bibinfo  {journal} {Phys. Rev. D}\ }\textbf {\bibinfo {volume}
  {94}},\ \bibinfo {pages} {083002} (\bibinfo {year} {2016})},\ \Eprint
  {http://arxiv.org/abs/1402.6671} {arXiv:1402.6671 [hep-ph]} \BibitemShut
  {NoStop}%
\bibitem [{\citenamefont {Dienes}\ \emph {et~al.}(2015)\citenamefont {Dienes},
  \citenamefont {Kumar}, \citenamefont {Thomas},\ and\ \citenamefont
  {Yaylali}}]{Dienes:2014via}%
  \BibitemOpen
  \bibfield  {author} {\bibinfo {author} {\bibfnamefont {K.~R.}\ \bibnamefont
  {Dienes}}, \bibinfo {author} {\bibfnamefont {J.}~\bibnamefont {Kumar}},
  \bibinfo {author} {\bibfnamefont {B.}~\bibnamefont {Thomas}}, \ and\ \bibinfo
  {author} {\bibfnamefont {D.}~\bibnamefont {Yaylali}},\ }\href {\doibase
  10.1103/PhysRevLett.114.051301} {\bibfield  {journal} {\bibinfo  {journal}
  {Phys. Rev. Lett.}\ }\textbf {\bibinfo {volume} {114}},\ \bibinfo {pages}
  {051301} (\bibinfo {year} {2015})},\ \Eprint {http://arxiv.org/abs/1406.4868}
  {arXiv:1406.4868 [hep-ph]} \BibitemShut {NoStop}%
\bibitem [{\citenamefont {Barello}\ \emph {et~al.}(2014)\citenamefont
  {Barello}, \citenamefont {Chang},\ and\ \citenamefont
  {Newby}}]{Barello:2014uda}%
  \BibitemOpen
  \bibfield  {author} {\bibinfo {author} {\bibfnamefont {G.}~\bibnamefont
  {Barello}}, \bibinfo {author} {\bibfnamefont {S.}~\bibnamefont {Chang}}, \
  and\ \bibinfo {author} {\bibfnamefont {C.~A.}\ \bibnamefont {Newby}},\ }\href
  {\doibase 10.1103/PhysRevD.90.094027} {\bibfield  {journal} {\bibinfo
  {journal} {Phys. Rev. D}\ }\textbf {\bibinfo {volume} {90}},\ \bibinfo
  {pages} {094027} (\bibinfo {year} {2014})},\ \Eprint
  {http://arxiv.org/abs/1409.0536} {arXiv:1409.0536 [hep-ph]} \BibitemShut
  {NoStop}%
\bibitem [{\citenamefont {Bramante}\ \emph {et~al.}(2016)\citenamefont
  {Bramante}, \citenamefont {Fox}, \citenamefont {Kribs},\ and\ \citenamefont
  {Martin}}]{Bramante:2016rdh}%
  \BibitemOpen
  \bibfield  {author} {\bibinfo {author} {\bibfnamefont {J.}~\bibnamefont
  {Bramante}}, \bibinfo {author} {\bibfnamefont {P.~J.}\ \bibnamefont {Fox}},
  \bibinfo {author} {\bibfnamefont {G.~D.}\ \bibnamefont {Kribs}}, \ and\
  \bibinfo {author} {\bibfnamefont {A.}~\bibnamefont {Martin}},\ }\href
  {\doibase 10.1103/PhysRevD.94.115026} {\bibfield  {journal} {\bibinfo
  {journal} {Phys. Rev. D}\ }\textbf {\bibinfo {volume} {94}},\ \bibinfo
  {pages} {115026} (\bibinfo {year} {2016})},\ \Eprint
  {http://arxiv.org/abs/1608.02662} {arXiv:1608.02662 [hep-ph]} \BibitemShut
  {NoStop}%
\bibitem [{\citenamefont {Bell}\ \emph {et~al.}(2018)\citenamefont {Bell},
  \citenamefont {Busoni},\ and\ \citenamefont {Robles}}]{Bell:2018pkk}%
  \BibitemOpen
  \bibfield  {author} {\bibinfo {author} {\bibfnamefont {N.~F.}\ \bibnamefont
  {Bell}}, \bibinfo {author} {\bibfnamefont {G.}~\bibnamefont {Busoni}}, \ and\
  \bibinfo {author} {\bibfnamefont {S.}~\bibnamefont {Robles}},\ }\href
  {\doibase 10.1088/1475-7516/2018/09/018} {\bibfield  {journal} {\bibinfo
  {journal} {JCAP}\ }\textbf {\bibinfo {volume} {09}},\ \bibinfo {pages} {018}
  (\bibinfo {year} {2018})},\ \Eprint {http://arxiv.org/abs/1807.02840}
  {arXiv:1807.02840 [hep-ph]} \BibitemShut {NoStop}%
\bibitem [{\citenamefont {Jordan}\ \emph {et~al.}(2018)\citenamefont {Jordan},
  \citenamefont {Kahn}, \citenamefont {Krnjaic}, \citenamefont {Moschella},\
  and\ \citenamefont {Spitz}}]{Jordan:2018gcd}%
  \BibitemOpen
  \bibfield  {author} {\bibinfo {author} {\bibfnamefont {J.~R.}\ \bibnamefont
  {Jordan}}, \bibinfo {author} {\bibfnamefont {Y.}~\bibnamefont {Kahn}},
  \bibinfo {author} {\bibfnamefont {G.}~\bibnamefont {Krnjaic}}, \bibinfo
  {author} {\bibfnamefont {M.}~\bibnamefont {Moschella}}, \ and\ \bibinfo
  {author} {\bibfnamefont {J.}~\bibnamefont {Spitz}},\ }\href {\doibase
  10.1103/PhysRevD.98.075020} {\bibfield  {journal} {\bibinfo  {journal} {Phys.
  Rev. D}\ }\textbf {\bibinfo {volume} {98}},\ \bibinfo {pages} {075020}
  (\bibinfo {year} {2018})},\ \Eprint {http://arxiv.org/abs/1806.05185}
  {arXiv:1806.05185 [hep-ph]} \BibitemShut {NoStop}%
\bibitem [{\citenamefont {Dutta}\ \emph {et~al.}(2019)\citenamefont {Dutta},
  \citenamefont {Ghosh},\ and\ \citenamefont {Kumar}}]{Dutta:2019fxn}%
  \BibitemOpen
  \bibfield  {author} {\bibinfo {author} {\bibfnamefont {B.}~\bibnamefont
  {Dutta}}, \bibinfo {author} {\bibfnamefont {S.}~\bibnamefont {Ghosh}}, \ and\
  \bibinfo {author} {\bibfnamefont {J.}~\bibnamefont {Kumar}},\ }\href
  {\doibase 10.1103/PhysRevD.100.075028} {\bibfield  {journal} {\bibinfo
  {journal} {Phys. Rev. D}\ }\textbf {\bibinfo {volume} {100}},\ \bibinfo
  {pages} {075028} (\bibinfo {year} {2019})},\ \Eprint
  {http://arxiv.org/abs/1905.02692} {arXiv:1905.02692 [hep-ph]} \BibitemShut
  {NoStop}%
\bibitem [{\citenamefont {Bell}\ \emph
  {et~al.}(2020{\natexlab{b}})\citenamefont {Bell}, \citenamefont {Dent},
  \citenamefont {Dutta}, \citenamefont {Ghosh}, \citenamefont {Kumar},\ and\
  \citenamefont {Newstead}}]{Bell:2020bes}%
  \BibitemOpen
  \bibfield  {author} {\bibinfo {author} {\bibfnamefont {N.~F.}\ \bibnamefont
  {Bell}}, \bibinfo {author} {\bibfnamefont {J.~B.}\ \bibnamefont {Dent}},
  \bibinfo {author} {\bibfnamefont {B.}~\bibnamefont {Dutta}}, \bibinfo
  {author} {\bibfnamefont {S.}~\bibnamefont {Ghosh}}, \bibinfo {author}
  {\bibfnamefont {J.}~\bibnamefont {Kumar}}, \ and\ \bibinfo {author}
  {\bibfnamefont {J.~L.}\ \bibnamefont {Newstead}},\ }\href {\doibase
  10.1103/PhysRevLett.125.161803} {\bibfield  {journal} {\bibinfo  {journal}
  {Phys. Rev. Lett.}\ }\textbf {\bibinfo {volume} {125}},\ \bibinfo {pages}
  {161803} (\bibinfo {year} {2020}{\natexlab{b}})},\ \Eprint
  {http://arxiv.org/abs/2006.12461} {arXiv:2006.12461 [hep-ph]} \BibitemShut
  {NoStop}%
\bibitem [{\citenamefont {Batell}\ \emph {et~al.}(2021)\citenamefont {Batell},
  \citenamefont {Berger}, \citenamefont {Darm\'e},\ and\ \citenamefont
  {Frugiuele}}]{Batell:2021ooj}%
  \BibitemOpen
  \bibfield  {author} {\bibinfo {author} {\bibfnamefont {B.}~\bibnamefont
  {Batell}}, \bibinfo {author} {\bibfnamefont {J.}~\bibnamefont {Berger}},
  \bibinfo {author} {\bibfnamefont {L.}~\bibnamefont {Darm\'e}}, \ and\
  \bibinfo {author} {\bibfnamefont {C.}~\bibnamefont {Frugiuele}},\ }\href@noop
  {} {\  (\bibinfo {year} {2021})},\ \Eprint {http://arxiv.org/abs/2106.04584}
  {arXiv:2106.04584 [hep-ph]} \BibitemShut {NoStop}%
\bibitem [{\citenamefont {Cappiello}\ \emph {et~al.}(2019)\citenamefont
  {Cappiello}, \citenamefont {Ng},\ and\ \citenamefont
  {Beacom}}]{Cappiello:2018hsu}%
  \BibitemOpen
  \bibfield  {author} {\bibinfo {author} {\bibfnamefont {C.~V.}\ \bibnamefont
  {Cappiello}}, \bibinfo {author} {\bibfnamefont {K.~C.~Y.}\ \bibnamefont
  {Ng}}, \ and\ \bibinfo {author} {\bibfnamefont {J.~F.}\ \bibnamefont
  {Beacom}},\ }\href {\doibase 10.1103/PhysRevD.99.063004} {\bibfield
  {journal} {\bibinfo  {journal} {Phys. Rev. D}\ }\textbf {\bibinfo {volume}
  {99}},\ \bibinfo {pages} {063004} (\bibinfo {year} {2019})},\ \Eprint
  {http://arxiv.org/abs/1810.07705} {arXiv:1810.07705 [hep-ph]} \BibitemShut
  {NoStop}%
\bibitem [{\citenamefont {Boschini}\ \emph {et~al.}(2017)\citenamefont
  {Boschini} \emph {et~al.}}]{Boschini:2017fxq}%
  \BibitemOpen
  \bibfield  {author} {\bibinfo {author} {\bibfnamefont {M.~J.}\ \bibnamefont
  {Boschini}} \emph {et~al.},\ }\href {\doibase 10.3847/1538-4357/aa6e4f}
  {\bibfield  {journal} {\bibinfo  {journal} {Astrophys. J.}\ }\textbf
  {\bibinfo {volume} {840}},\ \bibinfo {pages} {115} (\bibinfo {year}
  {2017})},\ \Eprint {http://arxiv.org/abs/1704.06337} {arXiv:1704.06337
  [astro-ph.HE]} \BibitemShut {NoStop}%
\bibitem [{\citenamefont {Boddy}\ \emph {et~al.}(2017)\citenamefont {Boddy},
  \citenamefont {Dienes}, \citenamefont {Kim}, \citenamefont {Kumar},
  \citenamefont {Park},\ and\ \citenamefont {Thomas}}]{Boddy:2016hbp}%
  \BibitemOpen
  \bibfield  {author} {\bibinfo {author} {\bibfnamefont {K.~K.}\ \bibnamefont
  {Boddy}}, \bibinfo {author} {\bibfnamefont {K.~R.}\ \bibnamefont {Dienes}},
  \bibinfo {author} {\bibfnamefont {D.}~\bibnamefont {Kim}}, \bibinfo {author}
  {\bibfnamefont {J.}~\bibnamefont {Kumar}}, \bibinfo {author} {\bibfnamefont
  {J.-C.}\ \bibnamefont {Park}}, \ and\ \bibinfo {author} {\bibfnamefont
  {B.}~\bibnamefont {Thomas}},\ }\href {\doibase 10.1103/PhysRevD.95.055024}
  {\bibfield  {journal} {\bibinfo  {journal} {Phys. Rev. D}\ }\textbf {\bibinfo
  {volume} {95}},\ \bibinfo {pages} {055024} (\bibinfo {year} {2017})},\
  \Eprint {http://arxiv.org/abs/1609.09104} {arXiv:1609.09104 [hep-ph]}
  \BibitemShut {NoStop}%
\bibitem [{\citenamefont {Helm}(1956)}]{Helm:1956zz}%
  \BibitemOpen
  \bibfield  {author} {\bibinfo {author} {\bibfnamefont {R.~H.}\ \bibnamefont
  {Helm}},\ }\href {\doibase 10.1103/PhysRev.104.1466} {\bibfield  {journal}
  {\bibinfo  {journal} {Phys. Rev.}\ }\textbf {\bibinfo {volume} {104}},\
  \bibinfo {pages} {1466} (\bibinfo {year} {1956})}\BibitemShut {NoStop}%
\bibitem [{\citenamefont {Dienes}\ \emph {et~al.}(2017)\citenamefont {Dienes},
  \citenamefont {Kumar}, \citenamefont {Thomas},\ and\ \citenamefont
  {Yaylali}}]{Dienes:2017ylr}%
  \BibitemOpen
  \bibfield  {author} {\bibinfo {author} {\bibfnamefont {K.~R.}\ \bibnamefont
  {Dienes}}, \bibinfo {author} {\bibfnamefont {J.}~\bibnamefont {Kumar}},
  \bibinfo {author} {\bibfnamefont {B.}~\bibnamefont {Thomas}}, \ and\ \bibinfo
  {author} {\bibfnamefont {D.}~\bibnamefont {Yaylali}},\ }\href {\doibase
  10.1103/PhysRevD.96.115009} {\bibfield  {journal} {\bibinfo  {journal} {Phys.
  Rev. D}\ }\textbf {\bibinfo {volume} {96}},\ \bibinfo {pages} {115009}
  (\bibinfo {year} {2017})},\ \Eprint {http://arxiv.org/abs/1708.09698}
  {arXiv:1708.09698 [hep-ph]} \BibitemShut {NoStop}%
\bibitem [{\citenamefont {Aprile}\ \emph {et~al.}(2018)\citenamefont {Aprile}
  \emph {et~al.}}]{XENON:2018voc}%
  \BibitemOpen
  \bibfield  {author} {\bibinfo {author} {\bibfnamefont {E.}~\bibnamefont
  {Aprile}} \emph {et~al.} (\bibinfo {collaboration} {XENON}),\ }\href
  {\doibase 10.1103/PhysRevLett.121.111302} {\bibfield  {journal} {\bibinfo
  {journal} {Phys. Rev. Lett.}\ }\textbf {\bibinfo {volume} {121}},\ \bibinfo
  {pages} {111302} (\bibinfo {year} {2018})},\ \Eprint
  {http://arxiv.org/abs/1805.12562} {arXiv:1805.12562 [astro-ph.CO]}
  \BibitemShut {NoStop}%
\bibitem [{\citenamefont {Nussinov}\ \emph {et~al.}(2009)\citenamefont
  {Nussinov}, \citenamefont {Wang},\ and\ \citenamefont
  {Yavin}}]{Nussinov:2009ft}%
  \BibitemOpen
  \bibfield  {author} {\bibinfo {author} {\bibfnamefont {S.}~\bibnamefont
  {Nussinov}}, \bibinfo {author} {\bibfnamefont {L.-T.}\ \bibnamefont {Wang}},
  \ and\ \bibinfo {author} {\bibfnamefont {I.}~\bibnamefont {Yavin}},\ }\href
  {\doibase 10.1088/1475-7516/2009/08/037} {\bibfield  {journal} {\bibinfo
  {journal} {JCAP}\ }\textbf {\bibinfo {volume} {08}},\ \bibinfo {pages} {037}
  (\bibinfo {year} {2009})},\ \Eprint {http://arxiv.org/abs/0905.1333}
  {arXiv:0905.1333 [hep-ph]} \BibitemShut {NoStop}%
\bibitem [{\citenamefont {Menon}\ \emph {et~al.}(2010)\citenamefont {Menon},
  \citenamefont {Morris}, \citenamefont {Pierce},\ and\ \citenamefont
  {Weiner}}]{Menon:2009qj}%
  \BibitemOpen
  \bibfield  {author} {\bibinfo {author} {\bibfnamefont {A.}~\bibnamefont
  {Menon}}, \bibinfo {author} {\bibfnamefont {R.}~\bibnamefont {Morris}},
  \bibinfo {author} {\bibfnamefont {A.}~\bibnamefont {Pierce}}, \ and\ \bibinfo
  {author} {\bibfnamefont {N.}~\bibnamefont {Weiner}},\ }\href {\doibase
  10.1103/PhysRevD.82.015011} {\bibfield  {journal} {\bibinfo  {journal} {Phys.
  Rev. D}\ }\textbf {\bibinfo {volume} {82}},\ \bibinfo {pages} {015011}
  (\bibinfo {year} {2010})},\ \Eprint {http://arxiv.org/abs/0905.1847}
  {arXiv:0905.1847 [hep-ph]} \BibitemShut {NoStop}%
\bibitem [{\citenamefont {Kumar}\ \emph {et~al.}(2012)\citenamefont {Kumar},
  \citenamefont {Learned}, \citenamefont {Smith},\ and\ \citenamefont
  {Richardson}}]{Kumar:2012uh}%
  \BibitemOpen
  \bibfield  {author} {\bibinfo {author} {\bibfnamefont {J.}~\bibnamefont
  {Kumar}}, \bibinfo {author} {\bibfnamefont {J.~G.}\ \bibnamefont {Learned}},
  \bibinfo {author} {\bibfnamefont {S.}~\bibnamefont {Smith}}, \ and\ \bibinfo
  {author} {\bibfnamefont {K.}~\bibnamefont {Richardson}},\ }\href {\doibase
  10.1103/PhysRevD.86.073002} {\bibfield  {journal} {\bibinfo  {journal} {Phys.
  Rev. D}\ }\textbf {\bibinfo {volume} {86}},\ \bibinfo {pages} {073002}
  (\bibinfo {year} {2012})},\ \Eprint {http://arxiv.org/abs/1204.5120}
  {arXiv:1204.5120 [hep-ph]} \BibitemShut {NoStop}%
\bibitem [{\citenamefont {Abud}\ \emph {et~al.}(2021)\citenamefont {Abud} \emph
  {et~al.}}]{DUNE:2021gbm}%
  \BibitemOpen
  \bibfield  {author} {\bibinfo {author} {\bibfnamefont {A.~A.}\ \bibnamefont
  {Abud}} \emph {et~al.} (\bibinfo {collaboration} {DUNE}),\ }\href@noop {} {\
  (\bibinfo {year} {2021})},\ \Eprint {http://arxiv.org/abs/2107.09109}
  {arXiv:2107.09109 [hep-ex]} \BibitemShut {NoStop}%
\bibitem [{\citenamefont {Angloher}\ \emph {et~al.}(2017)\citenamefont
  {Angloher} \emph {et~al.}}]{CRESST:2017ues}%
  \BibitemOpen
  \bibfield  {author} {\bibinfo {author} {\bibfnamefont {G.}~\bibnamefont
  {Angloher}} \emph {et~al.} (\bibinfo {collaboration} {CRESST}),\ }\href
  {\doibase 10.1140/epjc/s10052-017-5223-9} {\bibfield  {journal} {\bibinfo
  {journal} {Eur. Phys. J. C}\ }\textbf {\bibinfo {volume} {77}},\ \bibinfo
  {pages} {637} (\bibinfo {year} {2017})},\ \Eprint
  {http://arxiv.org/abs/1707.06749} {arXiv:1707.06749 [astro-ph.CO]}
  \BibitemShut {NoStop}%
\bibitem [{\citenamefont {Erickcek}\ \emph {et~al.}(2007)\citenamefont
  {Erickcek}, \citenamefont {Steinhardt}, \citenamefont {McCammon},\ and\
  \citenamefont {McGuire}}]{Erickcek:2007jv}%
  \BibitemOpen
  \bibfield  {author} {\bibinfo {author} {\bibfnamefont {A.~L.}\ \bibnamefont
  {Erickcek}}, \bibinfo {author} {\bibfnamefont {P.~J.}\ \bibnamefont
  {Steinhardt}}, \bibinfo {author} {\bibfnamefont {D.}~\bibnamefont
  {McCammon}}, \ and\ \bibinfo {author} {\bibfnamefont {P.~C.}\ \bibnamefont
  {McGuire}},\ }\href {\doibase 10.1103/PhysRevD.76.042007} {\bibfield
  {journal} {\bibinfo  {journal} {Phys. Rev. D}\ }\textbf {\bibinfo {volume}
  {76}},\ \bibinfo {pages} {042007} (\bibinfo {year} {2007})},\ \Eprint
  {http://arxiv.org/abs/0704.0794} {arXiv:0704.0794 [astro-ph]} \BibitemShut
  {NoStop}%
\bibitem [{\citenamefont {Boddy}\ and\ \citenamefont
  {Gluscevic}(2018)}]{Boddy:2018kfv}%
  \BibitemOpen
  \bibfield  {author} {\bibinfo {author} {\bibfnamefont {K.~K.}\ \bibnamefont
  {Boddy}}\ and\ \bibinfo {author} {\bibfnamefont {V.}~\bibnamefont
  {Gluscevic}},\ }\href {\doibase 10.1103/PhysRevD.98.083510} {\bibfield
  {journal} {\bibinfo  {journal} {Phys. Rev. D}\ }\textbf {\bibinfo {volume}
  {98}},\ \bibinfo {pages} {083510} (\bibinfo {year} {2018})},\ \Eprint
  {http://arxiv.org/abs/1801.08609} {arXiv:1801.08609 [astro-ph.CO]}
  \BibitemShut {NoStop}%
\bibitem [{\citenamefont {Ackermann}\ \emph {et~al.}(2015)\citenamefont
  {Ackermann} \emph {et~al.}}]{Fermi-LAT:2015qzw}%
  \BibitemOpen
  \bibfield  {author} {\bibinfo {author} {\bibfnamefont {M.}~\bibnamefont
  {Ackermann}} \emph {et~al.} (\bibinfo {collaboration} {Fermi-LAT}),\ }\href
  {\doibase 10.1088/1475-7516/2015/09/008} {\bibfield  {journal} {\bibinfo
  {journal} {JCAP}\ }\textbf {\bibinfo {volume} {09}},\ \bibinfo {pages} {008}
  (\bibinfo {year} {2015})},\ \Eprint {http://arxiv.org/abs/1501.05464}
  {arXiv:1501.05464 [astro-ph.CO]} \BibitemShut {NoStop}%
\end{thebibliography}%

\end{document}